\documentclass[%
reprint,
amsmath,amssymb,
aps,
pra,
floatfix,
]{revtex4-2}

\usepackage{graphicx}
\usepackage{dcolumn}
\usepackage{bm}
\usepackage{subfigure}
\usepackage{diagbox}

\begin{document}
	\preprint{APS/123-QED}
	
	\title{Heralded qudit-based high-dimensional entanglement generation for hybrid photon-emitter system by waveguide-mediated scattering}
	

	
	

\author{Fang-Fang Du$^1$} \email{Duff@nuc.edu.cn}
\author{Ling-Hui Li$^1$}
\author{Qiu-Lin Tan$^1$}
\author{Zhuo-Ya Bai$^2,$}
\email{zybai@mail.tsinghua.edu.cn}
\affiliation{$^1$Key Laboratory of Micro/nano Devices and Systems, Ministry of Education, North University of China, Tai Yuan 030051, China}

\affiliation{$^2$Beijing National Research Center for Information Science and Technology, Department of Electronic Engineering, Tsinghua University, Beijing 100084, China}
	
	\date{\today}
	
	\begin{abstract}
		Quantum entanglement systems based on qudits dilate high-dimensional (HD) state spaces and enhance resistance to loss in quantum information processing (QIP). To fully exploit this potential, effective schemes for generating HD entanglement are crucial.
		In this paper, we propose a flexible heralded scheme generating random 4D two-qudit maximal entanglement for hybrid photon-emitter system by
		entering different input ports.
		This approach can be further extended to prepare 4D $n$-qudit ($n \geq 3$) maximal entanglement utilizing the 4D single-qudit $Z^{m} (m=1,2,3)$ gate for the first qudit and $X^{m}$ gate for the other qudits (except the second qudit). For the hybrid system, the first 4D qudit is encoded on the hybrid polarization-path states of a flying photon, while the second and subsequent 4D qudits are represented by two stationary emitters coupled to respective 1D waveguide.
		The qudit-encoded hybrid  HD entanglement offers advantages over economizing quantum resource without any auxiliary qudits, and
		obtaining robust fidelities of various HD entanglement by the error-detected mechanism
		of the emitter-waveguide systems. Moreover, the proposed protocol can be spread to generate $d$D
		$n$-qudit ($d\geq 2^{p+1}$, $n, p=2,3,\dots$) entangled states, further broadening its applicability in HD QIP.
	\end{abstract}
	
	
	\maketitle
	
	\section{Introduction} \label{sec1}
	Quantum entanglement, manifesting the intrinsic nonlocality of quantum system, not only reveals the profound differences between the quantum and classics, but also provides the way for extensive applications in the realm of quantum information processing (QIP)
	\cite{sx1,nielsen2010quantum,lodahl2017chiral}, encompassing quantum key distribution \cite{xq35,Qkd,shengQKD2}, quantum teleportation  \cite{sx6,sx7,QT}, ultrafast quantum secure direct communication \cite{shengQSDC1,shengQSDC2,shengQSDC3,zhang2022realization}, and quantum secret sharing \cite{shengQSS1,shengQSS2}. A common illustration of quantum entanglement involves 2D bipartite systems \cite{MultiparticleEg1,MultiparticleEg2,MultiparticleEg3,One-WayQC,PhysRevA.62.062314}. Whereas, quantum systems can also exhibit more intricate entanglement structures. Recent research has expanded to explore these richer forms of entanglement, including cases where entanglement arises simultaneously across multiple properties of a photonic system—a phenomenon referred to as hyperentanglement \cite{high12,high13,OAM,spatial}, as well as qudit-encoded high-dimensional (HD) entanglement. 
	In particular, HD entanglement enhances quantum transmission capacities and potential for practical applications, enabling dense coding and offering additional benefits comprising improved security in communication systems \cite{high1,xq-APL,high3,high4,high5,du2025deterministic}. As a result, the practical realization of HD entangled states in $d$D ($d \geq 3$) qudit-encoded systems opens new opportunities for advanced and powerful quantum technologies \cite{Qkd2,Experimental4p,high-dimensional,PhysRevApplied.14.054057,PRA062101,krenn2014generation,malik2016multi,schaeff2015experimental}.
	
	Photons offer numerous advantages in quantum entanglement research, taking an edge in terms of high transmission speed, precise information processing, and strong resistance to decoherence. Beyond polarization as a DoF for encoding information, photons can also utilize other DoFs, such as time, frequency, orbital angular momentum, and path modes, all of which enable the encoding of HD qudit \cite{102,highOAM,PRA052301}.
	In the realm of linear optics, there are two primary approaches for generating HD entangled states. One method involves the superposition of spatial modes to produce a $d$D $N$-photon entangled state with an efficiency of $d(2/Nd)^{N/2}$ \cite{krenn2017entanglement,reimer2019high}. The other approach is based on Fourier transform matrices in linear optical circuits to construct HD entangled states \cite{xq42,LPR-high2}. However, the efficiency of the latter decreases as the dimension and photon count increase. Due to the nondeterministic nature of linear optical systems, achieving deterministic HD entangled states suitable for QIP requires exploiting the interaction between photon and matter medium.
	
	Waveguide cavity quantum electrodynamics (QED) has emerged as a promising platform for scalable photonic quantum computing, while also enabling non-destructive and high-precision quantum state detection. This theory describes the interaction between proximal quantum emitters and 1D  waveguide-trapped electromagnetic modes \cite{song65,song75,song82,PhysRevA.68.013803,PhysRevLett.101.100501,song2017photon,PRL213001,PRL153601}. Waveguides enable key functionalities in diverse quantum platforms, including photonic crystals, diamond-based circuits, superconducting microwave resonators, and nanofiber optical systems. In 2005,  Shen et al.\cite{song65} put forward a novel proposal for realizing the coupling between a single quantum emitter embedded in 1D waveguide and a photon, which can be considered as a "bad" cavity. More recently, there has been increasing focus on studying photon transmission within 1D waveguide-emitter system, with comprehensive theoretical and experimental studies currently in progress. Song et al. \cite{song} investigated the quantum scattering phenomena between a weak coherent optical field and an ensemble of $\rm{\Lambda}$-type emitters coupled to a 1D waveguide, subsequently developing a new heralded quantum repeater concept  and developed a framework for universal quantum gates and entanglement generation for emitter-waveguide systems.\cite{song2021heralded,song-creation}

	In the paper, by using the scattering properties of a single photon interacting with a single emitter
	in 1D waveguide, we first propose a scheme for generating random 4D two-qudit entanglement using two photon-emitter systems, which can be applied to the generation of 4D $n$-qudit ($n\geq3$) entangled states. Additionally, we utilize linear optical components only to implement 4D Pauli $Z$ gates and utilize two  error-detected emitter-waveguide
	unions  to implement 4D Pauli $X$ gates, which can be used to generate arbitrary 4D qudit-encoded $n$-qudit entangled states. Moreover, the protocol can also be extended to the generation of $d$D
	$n$-qudit ($d\geq 2^{p+1}$, $n, p=2,3,\dots$) entangled states. Furthermore, by the error-detected mechanism
	of 1D waveguide-emitter system, the generation protocols of HD entangled states are provided with high fidelities and efficiencies.
	
	The organization of the paper is as follows. An error-detected emitter-waveguide
	union is addressed briefly in Sec. \ref{sec2}. In Sec. \ref{sec3}, the 4D two-qudit and three-qudit entanglement generation for hybrid system is presented. Besides, 4D single-qudit $Z$ gates implemented via linear optics components are introduced for obtaining arbitrary 4D two-qudit entangled states. For generating three-qudit entanglement, 4D single-qudit $X$ gates implemented via linear optics components and the error-detected emitter-waveguide unions are introduced for obtaining arbitrary 4D three-qudit entangled states.  In Sec. \ref{sec4}, the generation of 4D $n$-qudit entangled states is presented. Entanglement preparation for two-qudit 8D entangled states are extended in Sec. \ref{sec5}.  Fidelities and Efficiencies of our protocols are discussed in Sec. \ref{sec6}. At last, the summary is presented in Sec. \ref{sec7}.

	\section{The 1D waveguide-emitter system}\label{sec2}
	\begin{figure}[htbp]
		\begin{center}	\includegraphics[width=8 cm,angle=0]{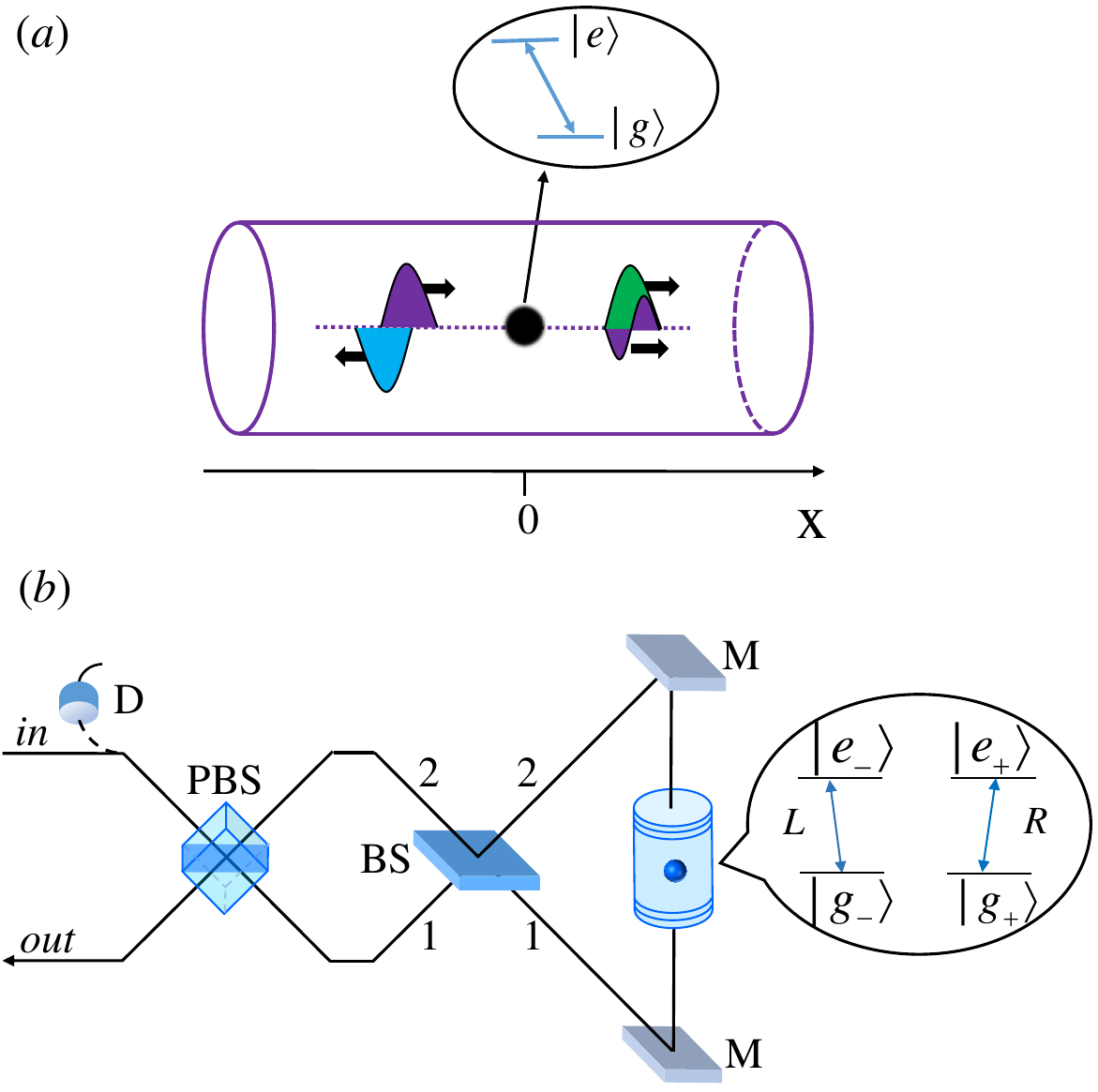}	\caption{(a) Schematic representation of a two-level  emitter (black circular dot) coupled within a 1D waveguide (cylindrical structure), where the emitter located at $x=0$ possesses ground state $|g\rangle$  and excited state $|e\rangle$. (b) Error-detected emitter-waveguide union for enhancing photon scattering with a four-level emitter in the waveguide, incorporating a balanced beam splitter (BS) and polarizing beam splitter (PBS) that transmits $\vert H \rangle$ and reflects $\vert V \rangle$ photon states. D represent a single photon detector.} \label{fig1}	
		\end{center}
	\end{figure}
	The setup shown in Fig. \ref{fig1}(a) illustrates a quantum system in which a two-level emitter interacts with the electromagnetic modes of a 1D waveguide. The emitter, characterized by its ground state $|g\rangle$ and excited state $|e\rangle$, has an energy difference of $\hbar \omega_{a}$. The interaction between the two-level emitter and the waveguide is described by the Jaynes-Cummings model. The corresponding Hamiltonian is expressed as \cite{chang2007single}
	\begin{eqnarray} \label{eq1}
		H &=& \hbar \omega_{a} \sigma_{+}\sigma_{-} + \sum_{k} \hbar \omega_{k} a_{k}^{\dagger}a_{k} \nonumber\\
		&&+ \sum_{k}\left( g_{k}\sigma_{+}a_{k} + g_{k}^{*}\sigma_{-}a_{k}^{\dagger}\right).
	\end{eqnarray}
	Here, $\sigma_{+}$ and $\sigma_{-}$ correspond to the emitter's raising and lowering operators, while $\omega_{k}$ denotes the frequency of mode $k$. The bosonic field is described by annihilation and creation operators $a_{k}$ and $a_{k}^{\dagger}$, with $g_{k}$ characterizing the interaction strength between the emitter and mode $k$. Expressed in real-space coordinates as shown in Eq. \ref{eq2}, the Hamiltonian takes the form:
	\begin{eqnarray} \label{eq2}
		H'&=&\hbar\int dk\omega_{k}a_{k}^{\dag}a_{k}+\hbar g_{k}\int dk(a_{k}\sigma_{+}e^{ikx_{a}}\nonumber\\
		&&+h.c.)+\hbar(\omega_{a}-\frac{i\gamma'}{2})\sigma_{ee},
	\end{eqnarray}
	where $k$ represents the wave vector, $x_{a}$ denotes position of the atom, and $\omega_{k} = c|k|$. Besides, the decay rate $\gamma'$ accounts for free-space emission, typically arising from spontaneous emission processes. Assuming distinct left- and right-polarized fields, the field operator $a_{k}$ is decomposed as $a_{k} = a_{k,R} + a_{k,L}$. For a photon with energy $E_{K}$ incident from the left, the system state evolves to describe its scattering behavior, capturing interactions between the photon and the emitter. The system's quantum state is characterized by \cite{chang2007single}
	\begin{eqnarray} \label{eq3}
		\vert \psi_{k}\rangle&=&\int dx[\phi_{L}(x)\tau_{L}^{\dag}(x)+\phi_{R}(x)\tau_{R}^{\dag}(x)]\vert g,vac\rangle\nonumber\\
		&&+c_{e}\vert e,vac\rangle.
	\end{eqnarray}
	Here, $c_{e}$ represents the probability amplitude for the emitter to be in the excited state, while $\tau_{L}^{\dag}(x)$ and $\tau_{R}^{\dag}(x)$ denote bosonic operators that generate left- and right-polarized photons, respectively, at position $x$. The vacuum state of the photon field is represented by $\vert vac\rangle$. Additionally, the functions $\phi_{R}(x)$  ($\phi_{L}(x)$) describes the probability amplitudes of right-(left-) traveling photons,  and can be expressed as
	\begin{eqnarray} \label{eq4}
		\phi_{R}(x)&=&e^{ikx}\vartheta(-x)+te^{ikx}\vartheta(x),\nonumber\\
		\phi_{L}(x)&=&re^{-ikx}\vartheta(-x),
	\end{eqnarray}
	where $t$  ($r$) represents the transmission  (reflection) coefficient. Within the defined function framework, the Heaviside step function satisfies $\vartheta(x)=1$ for $x$ is greater than $0$ or $\vartheta(x)=0$ for $x$ is less than $0$. Through tackling the scattering eigenvalue equation $H\vert \psi_{k}\rangle=\psi_{k}\vert \psi_{k}\rangle$.
	Results include the reflection coefficient $ r $ and transmission coefficient $ t $ for the input photon, i.e.,
	\begin{eqnarray} \label{eq5}
		r=-\frac{1}{1+\gamma'/\gamma_{1D}-2i\Delta/\gamma_{1D}},\;\;
		t=r+1.
	\end{eqnarray}
	Here, $P = \gamma_{1D}/\gamma'$ measures waveguide emission enhancement, where $\gamma_{1D} = 4\pi g^2/c$ as the emitter's dissipation rate and $\Delta = \omega_k - \omega_a$ as the photon-emitter detuning. At resonance ($\Delta = 0$), the reflection coefficient is $r = -1/(1/P+1)$. For $P \gg 1$, the emitter serves as a mirror (i.e., $r \approx -1$) with a $\pi$ phase shift.
	
	Considering that a four-level emitter has degenerate two ground states $|\uparrow\rangle$, $|\downarrow\rangle$, and two excited states  $|\uparrow\downarrow\Uparrow\rangle$, $|\uparrow\downarrow\Downarrow\rangle$ (as shown Fig. \ref{fig1}(b)). A self-assembly InGaAs/GaAs quantum dot coupled within a 1D waveguide enables left- or right-circularly polarized photons ($|L\rangle$ or $|R\rangle$) to drive transitions \cite{gerardot2008optical,warburton2013single,PhysRevB.67.161306,PRL160504}: $|\downarrow\rangle \leftrightarrow |\uparrow\downarrow\Downarrow\rangle$ and $|\uparrow\rangle \leftrightarrow |\uparrow\downarrow\Uparrow\rangle$. The incident photon with wave function $|\psi\rangle$ interacts with the emitter, and scattering governs the system's evolution, that is,
	\begin{eqnarray} \label{eq6}
		&&\vert g_{+}\rangle\vert \psi\rangle\vert R\rangle\rightarrow\vert g_{+}\rangle\vert \phi_{r}\rangle\vert R\rangle+\vert g_{+}\rangle\vert \phi_{t}\rangle\vert R\rangle,\nonumber\\
		&&\vert g_{-}\rangle\vert \psi\rangle\vert L\rangle\rightarrow\vert g_{-}\rangle\vert \phi_{r}\rangle\vert L\rangle+\vert g_{-}\rangle\vert \phi_{t}\rangle\vert L\rangle,\nonumber\\
		&&\vert g_{-}\rangle\vert \psi\rangle\vert R\rangle\rightarrow\vert g_{-}\rangle\vert \psi\rangle\vert R\rangle,\nonumber\\
		&&\vert g_{+}\rangle\vert \psi\rangle\vert L\rangle\rightarrow\vert g_{+}\rangle\vert \psi\rangle\vert L\rangle,
	\end{eqnarray}
	where $\vert \phi_{t}\rangle = t\vert \psi\rangle$ and $\vert \phi_{r}\rangle = r\vert \psi\rangle$ are the transmitted and reflected photon states. The relation $t=1+r$ allows immediate calculation of $\vert \phi_{t}\rangle=(\vert \psi\rangle+\vert \phi\rangle)/2$ and $\vert \phi_{r}\rangle=(-\vert \psi\rangle+\vert \phi\rangle)/2$. For the input photon in the horizontal ($|H\rangle$) or vertical ($|V\rangle$) polarization, that is, $|H\rangle=\frac{1}{\sqrt{2}}(|R\rangle+|L\rangle)$,  $|V\rangle=\frac{1}{\sqrt{2}}(|R\rangle-|L\rangle)$. The processes transition to
	\begin{eqnarray} \label{eq7}
		&&\vert g_{+}\rangle\vert \psi\rangle\vert H\rangle\rightarrow\vert g_{+}\rangle\vert \phi_{r}\rangle\vert V \rangle+\vert g_{+}\rangle\vert \phi_{t}\rangle\vert H \rangle,\nonumber\\
		&&\vert g_{-}\rangle\vert \psi\rangle\vert H\rangle\rightarrow-\vert g_{-}\rangle\vert \phi_{r}\rangle\vert V \rangle+\vert g_{-}\rangle\vert \phi_{t}\rangle\vert H \rangle.
	\end{eqnarray}
	From Eq. (\ref{eq7}), $|H\rangle$- or $|V\rangle$-polarized photon can be obtained from an input $|H\rangle$ photon, regardless of the emitter state. On the other hand, when a $|V\rangle$-polarized photon interacts with an emitter in state $|g_{-}\rangle$, it acquires a $\pi$ phase shift dependent on state.
	
	Fig. \ref{fig1}(b) shows an error-detected emitter-waveguide union. Here the effect of the balanced beam splitter (BS) for a photon on the paths with the
	basis $\{|a_{1}\rangle,|a_{2}\rangle\}$ can be represented by the matrix
	\begin{eqnarray} \label{eq8}   
		{\frac{1}{\sqrt{2}}\left[ \begin{array}{cc}
				1 & 1 \\
				1 & -1 \end{array}
			\right ]}.
	\end{eqnarray}
	
	The input photon, prepared in $|\psi\rangle |H\rangle^1$ transmits through a polarized beam splitter (PBS), separating $|H\rangle$ state from $|V\rangle$ state. The photon is subsequently divided by a 50:50 BS  and interaction with the emitter, transforming the initial state $\vert \phi_{0}\rangle=|\psi\rangle|H\rangle^{1}\otimes|g_{\pm}\rangle$ to
	\begin{eqnarray} \label{eq9}
		\vert \phi_{1}\rangle &=& \frac{1}{\sqrt{2}}\vert g_{\pm}\rangle\vert \phi_{t}\rangle\vert H\rangle^{1}\pm\frac{1}{\sqrt{2}}\vert g_{\pm}\rangle\vert \phi_{r}\rangle\vert V\rangle^{1}\nonumber\\
		&&+\frac{1}{\sqrt{2}}\vert g_{\pm}\rangle\vert \phi_{t}\rangle\vert H\rangle^{2}
		\pm\frac{1}{\sqrt{2}}\vert g_{\pm}\rangle\vert \phi_{r}\rangle\vert V\rangle^{2},
	\end{eqnarray}
	where the superscript $i$ ($i = 1, 2, 3, 4$) identifies the photon's path. After undergoing reflection at mirror (M), the photon is redirected back to the BS, producing 
	\begin{eqnarray} \label{eq10}
		\vert \phi_{2}\rangle &=&
		t\vert g_{\pm}\rangle\vert \psi\rangle\vert H\rangle^{1}\pm r\vert g_{\pm}\rangle\vert \psi\rangle\vert V\rangle^{1}.
	\end{eqnarray}
	Quantum destructive interference leads to complete suppression of the photon in path 2, while the photon emerges exclusively in path 1, undergoing further polarization splitting by the PBS.  In evidence, $|H\rangle$-polarized photon signals failure through the detector triggered, while  $|V\rangle$-polarized photon moves to path 2 for the desired transformation
	\begin{eqnarray} \label{eq11}
		\vert g_{\pm}\rangle\vert \psi\rangle\vert H\rangle^{1}\rightarrow \pm r\vert g_{\pm}\rangle\vert \psi\rangle\vert V\rangle^{out}.
	\end{eqnarray}
Analogously, for a photon initially prepared in the state $\vert \psi\rangle\vert V\rangle$ that propagates through both the PBS and BS along path 2 before re-emerging from the BS through the same path, the scattering dynamics can be characterized by the following evolution after eliminating the unaffected polarization component of the output photon.
	\begin{eqnarray} \label{eq12}
		\vert g_{\pm}\rangle\vert \psi\rangle\vert V\rangle^{2}\rightarrow \pm r\vert g_{\pm}\rangle\vert \psi\rangle\vert H\rangle^{out}.
	\end{eqnarray}
	If the state of waveguide-emitter system is in superposition state, that is, $|\pm\rangle=\frac{1}{\sqrt{2}}(\vert g_{-}\rangle\pm\vert g_{+}\rangle)$, the quantum evolution induced by scattering follows
	\begin{eqnarray} \label{eq13}
		|\pm\rangle \vert \psi\rangle\vert H\rangle^{1}\rightarrow - r\vert \mp \rangle\vert \psi\rangle\vert V\rangle^{out},\nonumber\\
		|\pm\rangle \vert \psi\rangle\vert V\rangle^{2}\rightarrow - r\vert \mp \rangle\vert \psi\rangle\vert H\rangle^{out}.
	\end{eqnarray}
	Considering the input and output quantum states of the photon and  emitter in Eq. (\ref{eq13}), it's not difficult to notice that
	the polarization state of the photon and  the state of the emitter have undergone a reversal
	accompanied by the reflection coefficient $-r$.
	Obviously, the emitter-waveguide union composed of the PBS, BS, D, and the emitter-waveguide system has
	error-predicted mechanism, which is available for the generation of HD entangled state for hybrid system.
	
	\section{The 4D two-qudit and three-qudit entanglement generation for hybrid system}\label{sec3}
	
	\subsection{4D two-qudit entanglement generation for hybrid system}\label{sec3.1}
	\begin{figure}[htbp]
		\begin{center}	\includegraphics[width=8.5 cm,angle=0]{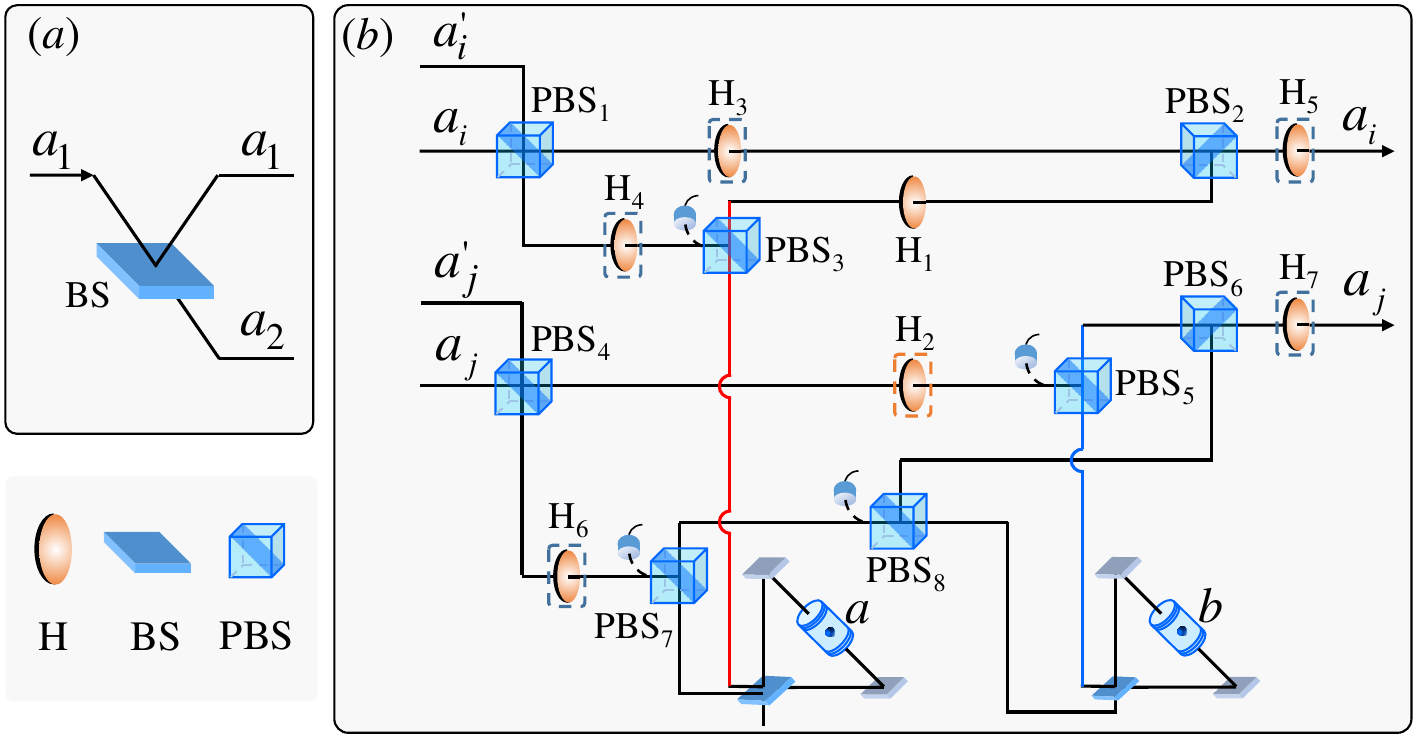}	\caption{(a) Schematic diagram for  initial state preparation of the single photon. (b) Schematic diagram for 4D two-qudit entanglement generation. H$_{i}$ ($i=1,2$) acts as the half-wave plate set at 45$^{\circ}$ that undergoes qubit-flip operation.}\label{fig2}
		\end{center}
	\end{figure}
	
	The two-qudit maximally entangled state, defined for qudit encoding with $d$D, can be expressed as \cite{gongshi}
	\begin{eqnarray}\label{eq14}
		|\varphi_{pq}\rangle=\frac1{\sqrt{d}}\sum_{l=0}^{d-1}e^{i2\pi lp/d}|l\rangle\otimes|l\oplus q\rangle_{12},
	\end{eqnarray}
	where $p$ and $q$ $(p, q = 0, 1, 2, ..., d - 1)$ represent the relative phase and spin of emitters information of $d$D entanglement, respectively. The subscripts 1 and 2 denote two entangled qudits. Besides, the relation $|l\rangle\otimes|l\oplus q\rangle=(l+q)$ shows modulo $d$.
	When $p = 0$, the 4D two-qudit maximally entangled states  can be written in the following forms
	\begin{eqnarray}\label{eq15}
		\left|\varphi_{00}\right\rangle &=& \frac12(|00\rangle+|11\rangle+|22\rangle+|33\rangle)_{12}, \nonumber\\
		\left|\varphi_{01}\right\rangle &=& \frac12(|01\rangle+|12\rangle+|23\rangle+|30\rangle)_{12},\nonumber\\
		\left|\varphi_{02}\right\rangle &=& \frac12(|02\rangle+|13\rangle+|20\rangle+|31\rangle)_{12},\nonumber\\
		\left|\varphi_{03}\right\rangle &=& \frac12(|03\rangle+|10\rangle+|21\rangle+|32\rangle)_{12}.
	\end{eqnarray}
	When $p = 1,2,3$, they can be expanded into
	\begin{eqnarray}\label{eq16}
		\left|\varphi_{1q}\right\rangle &=& \frac12(|0\rangle \otimes|0\oplus q\rangle+e^{i\frac{\pi}{2}}|1\rangle \otimes|1\oplus q\rangle\nonumber\\
		&&+e^{i\pi}|2\rangle \otimes|2\oplus q\rangle+e^{i\frac{3\pi}{2}}|3\rangle \otimes|3\oplus q\rangle)_{12}, \nonumber\\
		\left|\varphi_{2q}\right\rangle &=& \frac12(|0\rangle \otimes|0\oplus q\rangle+e^{i\pi}|1\rangle \otimes|1\oplus q\rangle\nonumber\\
		&&+|2\rangle \otimes|2\oplus q\rangle+e^{i\pi}|3\rangle \otimes|3\oplus q\rangle)_{12}, \nonumber\\
		\left|\varphi_{3q}\right\rangle &=& \frac12(|0\rangle \otimes|0\oplus q\rangle+e^{i\frac{3\pi}{2}}|1\rangle \otimes|1\oplus q\rangle\nonumber\\
		&&+e^{i\pi}|2\rangle \otimes|(2\oplus q\rangle+e^{i\frac{\pi}{2}}|3\rangle \otimes|3\oplus q\rangle)_{12}.
	\end{eqnarray}
	Specifically, the first 4D qudit is mapped to the hybrid polarization-path state of the flying photon and the second 4D qudit  is encoded on two stationary emitters $a$ and $b$, that is,
	\begin{eqnarray}\label{eq17}
		&&|Ha_1\rangle=|0\rangle_{1},\;\;| Ha_2\rangle=|1\rangle_{1},\nonumber\\
		&&|Va_1\rangle=|2\rangle_{1},\;\;| Va_2\rangle=|3\rangle_{1},\nonumber\\
		&&|++\rangle_{ab}=|0\rangle_{2},\;\;|+-\rangle_{ab}=|1\rangle_{2},\nonumber\\
		&&|-+\rangle_{ab}=|2\rangle_{2},\;\;|--\rangle_{ab}=|3\rangle_{2}.
	\end{eqnarray}
	Here, $|\pm\rangle=\frac{1}{\sqrt{2}}(\vert g_{-}\rangle\pm\vert g_{+}\rangle)$. 

	Fig. \ref{fig2} shows the schematic diagram for generating random 4D two-qudit maximally entangled state.
	Here, H$_{i}$ ($i=1,2$) represents the half-wave plate oriented at 45$^{\circ}$ that undergoes the qubit-flip operation, i.e.,
	$|H\rangle \leftrightarrow |V\rangle$.
	Supposed that the initial state of the photon is set in $|\Phi\rangle_{0}=\frac{1}{\sqrt{2}}(|H\rangle+|V\rangle)\otimes|a_{1}\rangle$.
	Firstly, the single photon goes through BS from path $a_{1}$,  as shown in Fig. \ref{fig2}(a), leading to
	\begin{eqnarray}\label{eq18}
		|\Phi\rangle_{1}=\frac{1}{2}(|H\rangle+|V\rangle)\otimes(|a_{1}\rangle+|a_{2}\rangle).
	\end{eqnarray}
	Since the single photon enters from different input ports to be used to generate diverse entangled states in Eq. (\ref{eq15}), the scheme is nimble. Here, two of four paths $a_{i}$, $a'_{i}$, $a_{j}$, and $a'_{j}$ ($i,j=1,2$) can connect with the above mentioned two paths $a_{1}$ and $a_{2}$
	shown in Tab. \ref{Table1}.
	As shown in Fig. \ref{fig2}(b),
	if we aim to achieve the entangled state $|\varphi_{00}\rangle$, two paths $a_{1}$ and $a_{2}$ should be connected to two input paths $a_{i} (i=1)$ and  $a_{j} (j=2)$, respectively.
	Immediately after, the  polarization photon $|H\rangle$ in path $a_{1}$ is directly transmitted by PBS$_{1}$ and PBS$_{2}$
	without interacting with the two emitters. The  polarization photon $|V\rangle$ in  path $a_{1}$ is reflected by PBS$_{1}$ and PBS$_{3}$,  and it interacts with emitter $a$ in the 1D waveguide, passes through PBS$_{3}$ and H$_{1}$  in sequence, and converges at PBS$_{2}$ of the path $a_{1}$.
	On the other hand, the $|H\rangle$-polarized photon in path $a_{2}$ undergoes a series of operations, i.e.,
	PBS$_{4}\rightarrow$ H$_{2}\rightarrow$ interaction with emitter $b\rightarrow$ PBS$_{5}\rightarrow$  PBS$_{6}$ in turn, meanwhile
	the $|V\rangle$-polarized photon in path $a_{2}$ undergoes a series of operations, i.e.,
	PBS$_{4}\rightarrow$PBS$_{7}\rightarrow$H$_{2}\rightarrow$ interaction with emitters $a$ and $b\rightarrow$   PBS$_{8}\rightarrow$  PBS$_{6}$.
	The state of the whole system changes from $|\Phi\rangle_{1} \otimes |+\rangle_a |+\rangle_b$ to $|\Phi\rangle_{2}$ where
	\begin{eqnarray}\label{eq19}
		|\Phi\rangle_{2}&=&	
		\frac{1}{2}(|Ha_1\rangle|+\rangle_a|+\rangle_b-r|Ha_2\rangle|+\rangle_a|-\rangle_b\nonumber\\
		&&-r|Va_1\rangle|-\rangle_a|+\rangle_b+r^{2}|Va_2\rangle|-\rangle_a|-\rangle_b)\nonumber\\
		&=&\frac{1}{2}(|00\rangle-r|11\rangle-r|22\rangle+r^{2}|33\rangle)_{12}.
	\end{eqnarray}
	To obtain the other three two-qudit entangled states with $p=0$ in Eq. (\ref{eq15}), the input two paths need to be adjusted, as shown in Tab. \ref{Table1}. In detail, to obtain the state $|\varphi_{01}\rangle$, two paths $a_{1}$ and $a_{2}$ should be connected to two input path $a_{j} (j=1)$ and  $a'_{i} (i=2)$, respectively. In addition, when the photon is input through path $a'_{i}$, it requires passing through three additional half-wave plates with $45^{\circ}$ (specifically H$_{3}$, H$_{4}$, and H$_{5}$), which are highlighted in Fig. \ref{fig2} with blue dashed boxes.
	The state of the whole system changes from $|\Phi\rangle_{1} \otimes |+\rangle_a |+\rangle_b$ to $|\Phi\rangle_{3}$, where
	\begin{eqnarray}\label{eq20}
		|\Phi\rangle_{3}&=&	
		\frac{1}{2}(-r|Ha_1\rangle|+\rangle_a|-\rangle_b-r|Ha_2\rangle|-\rangle_a|+\rangle_b\nonumber\\
		&&+r^{2}|Va_1\rangle|-\rangle_a|-\rangle_b+|Va_2\rangle|+\rangle_a|+\rangle_b)\nonumber\\
		&=&\frac{1}{2}(-r|01\rangle-r|12\rangle+r^{2}|23\rangle+|30\rangle)_{12}.
	\end{eqnarray}
	To obtain the state $|\varphi_{02}\rangle$, two paths $a_{1}$ and $a_{2}$ should be connected to two input path $a'_{i} (i=1)$ and  $a'_{j} (j=2)$, respectively. In addition, when the photon is input through path $a'_{j}$, it requires passing through two additional half-wave plates with $45^{\circ}$ (specifically H$_{6}$ and H$_{7}$), which are highlighted in Fig. \ref{fig2} with blue dashed boxes, while the wave plate H$_{2}$ that is explicitly marked by orange dashed boxes. The state $|\Phi\rangle_{4}$ can be obtained.
	\begin{eqnarray}\label{eq21}
		|\Phi\rangle_{4}&=&	
		\frac{1}{2}(-r|Ha_1\rangle|-\rangle_a|+\rangle_b+r^{2}|Ha_2\rangle|-\rangle_a|-\rangle_b\nonumber\\
		&&
		+|Va_1\rangle|+\rangle_a|+\rangle_b-r|Va_2\rangle|+\rangle_a|-\rangle_b)\nonumber\\
		&=&\frac{1}{2}(-r|02\rangle+r^{2}|13\rangle+|20\rangle-r|31\rangle)_{12}.
	\end{eqnarray}
	To obtain the state $|\varphi_{03}\rangle$, two paths $a_{1}$ and $a_{2}$ should be connected to two input path $a'_{j} (j=1)$ and  $a_{i} (i=2)$, respectively. Furthermore, half-wave plates H$_{6}$ and H$_{7}$ with $45^{\circ}$  must be included, with the explicit exclusion of H$_{2}$.  The state $|\Phi\rangle_{5}$ can be obtained.
	\begin{eqnarray}\label{eq22}
		|\Phi\rangle_{5}&=&	
		\frac{1}{2}(-r|Ha_1\rangle|-\rangle_a|-\rangle_b+r^{2}|Ha_2\rangle|+\rangle_a|+\rangle_b\nonumber\\
		&&
		+|Va_1\rangle|+\rangle_a|-\rangle_b-r|Va_2\rangle|-\rangle_a|+\rangle_b)\nonumber\\
		&=&\frac{1}{2}(r^{2}|03\rangle+|10\rangle-r|21\rangle-r|32\rangle)_{12}.
	\end{eqnarray}
Under the case of ideal scattering ($r = -1$), the above states $|\Phi\rangle_{2}$, $|\Phi\rangle_{3}$, $|\Phi\rangle_{4}$, and $|\Phi\rangle_{5}$
	are converted into the 4D two-qudit maximally entangled states $\left|\varphi_{00}\right\rangle$, $\left|\varphi_{01}\right\rangle$, $\left|\varphi_{02}\right\rangle$, and $\left|\varphi_{03}\right\rangle$ in Eq. (\ref{eq15}), respectively.
	\begin{figure}[htbp]
		\begin{center}	\includegraphics[width=8 cm,angle=0]{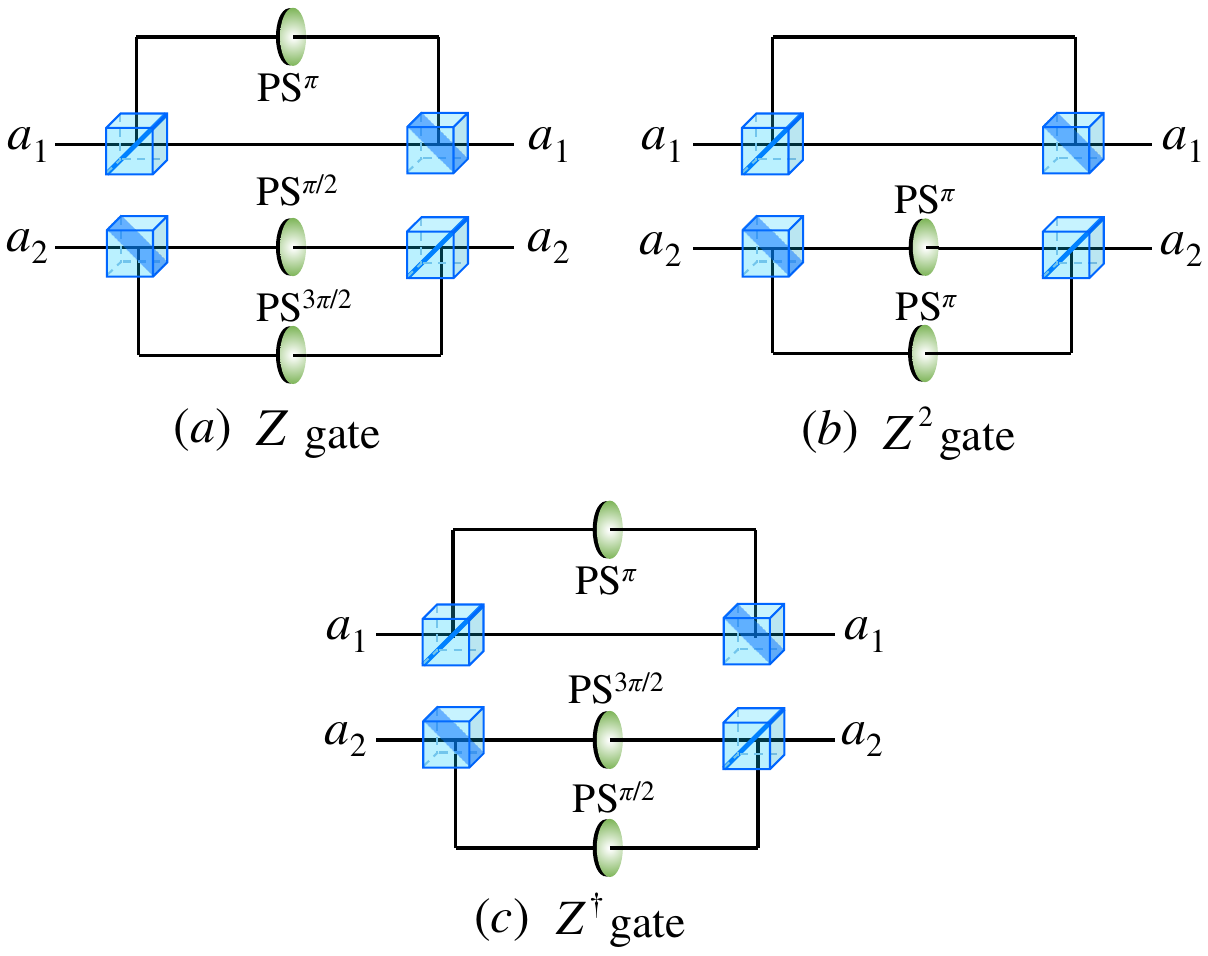}	\caption{The 4D $Z^{n}$ gates constructed by linear-optics elements. (a)$Z$ gate, (b)$Z^{2}$ gate, (c)$Z^{\dagger}$ gate.  PS$^{\theta}$ represents a phase shifter set at an angle $\theta$, introducing a mode-dependent phase factor of  $e^{i\theta}$ in the associated paths.} \label{fig3}	
		\end{center}
	\end{figure}
	
	The scheme is confined to producing only four particular two-qudit entangled states $|\varphi\rangle_{00}$, $|\varphi\rangle_{01}$, $|\varphi\rangle_{02}$, and $|\varphi\rangle_{03}$, which does not involve the generation of other entangled states in Eq. (\ref{eq16}). To enhance versatility in conveying a variety of information and broaden  utility in QIP tasks, mutual transformations between the 4D two-qudit entangled states need to introduce. The 4D single-qudit Pauli $Z^{m}$ gates, including $Z^{0} (I)$, $Z^{1} (Z)$, $Z^2$, and $Z^3$ ($Z^{\dagger}$) gates, as shown in Figs. \ref{fig3}$(a)$-$(c)$, are required, expressed by $Z^{m}|l\rangle=e^{2i\pi lm/4}|l\rangle$ ($m, l =0, 1, 2, 3$) \cite{LiuWQ}. These gates can transform the states $|\varphi\rangle_{00}$, $|\varphi\rangle_{01}$, $|\varphi\rangle_{02}$, and $|\varphi\rangle_{03}$ into any desired 4D two-qudit maximally entangled state.
	
	The matrix representations of four Pauli $Z^{m}$ ($m, l =0, 1, 2, 3$) gates for the single photon (the first qudit)
	with the specified basis $\{|0\rangle_{1}, |1\rangle_{1}, |2\rangle_{1}, |3\rangle_{1}\}$ are expressed as
	\begin{eqnarray}\label{eq23}
		Z^{0} &=& \left( \begin{array}{cccc}
			1 & 0 & 0 & 0 \\
			0& 1 & 0 & 0 \\
			0	&  0& 1 &0  \\
			0	& 0 & 0 & 1 \\
		\end{array} \right),\nonumber\\
		Z &=& \left( \begin{array}{cccc}
			1 & 0 & 0 & 0 \\
			0	& e^{i\frac{\pi}{2}} & 0 & 0 \\
			0	&  0& e^{i\pi} &0  \\
			0	&  0&  0& e^{i\frac{3\pi}{2}} \\
		\end{array} \right),\nonumber\\
		Z^{2} &=& \left( \begin{array}{cccc}
			1 & 0 &  0& 0 \\
			0	&  e^{i\pi} & 0 &0  \\
			0	& 0 & 1 &0  \\
			0	& 0 &  0& e^{i\pi} \\
		\end{array} \right),\nonumber
		\\ Z^{\dagger} &=& \left( \begin{array}{cccc}
			1 & 0 & 0 &0  \\
			0	& e^{i\frac{3\pi}{2}} &0  &0  \\
			0	& 0 & e^{i\pi} & 0 \\
			0	& 0 &  0& e^{i\frac{\pi}{2}} \\
		\end{array} \right).
	\end{eqnarray}
	Obviously, the two-qudit entangled state $|\varphi\rangle_{0q}$ can be transformed into $|\varphi\rangle_{1q}$, $|\varphi\rangle_{2q}$, and $|\varphi\rangle_{3q}$ by applying the Pauli $Z^{1}$ gate, $Z^2$ gate, and  $Z^3$ gate, respectively. We implement the different phase shifts by using the sandwich structure of the wave plates combination. Such as, $\pi/2$, $\pi$, and $3\pi/2$  phase shifts can be realized by the wave plates combination \cite{LiuWQ}: QWP$^{90^\circ}$-HWP$^{0^\circ}$-QWP$^{0^\circ}$, QWP$^{0^\circ}$-HWP$^{90^\circ}$-QWP$^{0^\circ}$, and QWP$^{90^\circ}$-HWP$^{90^\circ}$-QWP$^{0^\circ}$ in sequence.
	Therefore, random 4D two-qudit maximally entangled states for hybrid system can be realized within the HD state space.
	\begin{table}[htbp]
		\centering
		\caption{The relationship between four entangled states and input paths of  $a_{1}$ and $a_{2}$ for 4D 2-qudit entanglement generation.} \label{Table1}
		\begin{ruledtabular}
		\begin{tabular} {ccccc}
			& 	$\left|\varphi_{00}\right\rangle$ &	$\left|\varphi_{01}\right\rangle$ & 	$\left|\varphi_{02}\right\rangle$ &	$\left|\varphi_{03}\right\rangle$   \\
			\hline
			$a_{1}$ & $a_{i}$ & $a_{j}$ &$a'_{i}$ & $a'_{j}$\\
			$a_{2}$ & $a_{j}$ & $a'_{i}$ &$a'_{j}$ &$a_{i}$\\
	\end{tabular}
\end{ruledtabular}
	\end{table}
	
	\subsection{4D three-qudit entanglement generation for hybrid system}\label{sec3.2}
	
	\begin{figure*}[htbp]
		\begin{center}	\includegraphics[width=10cm,angle=0]{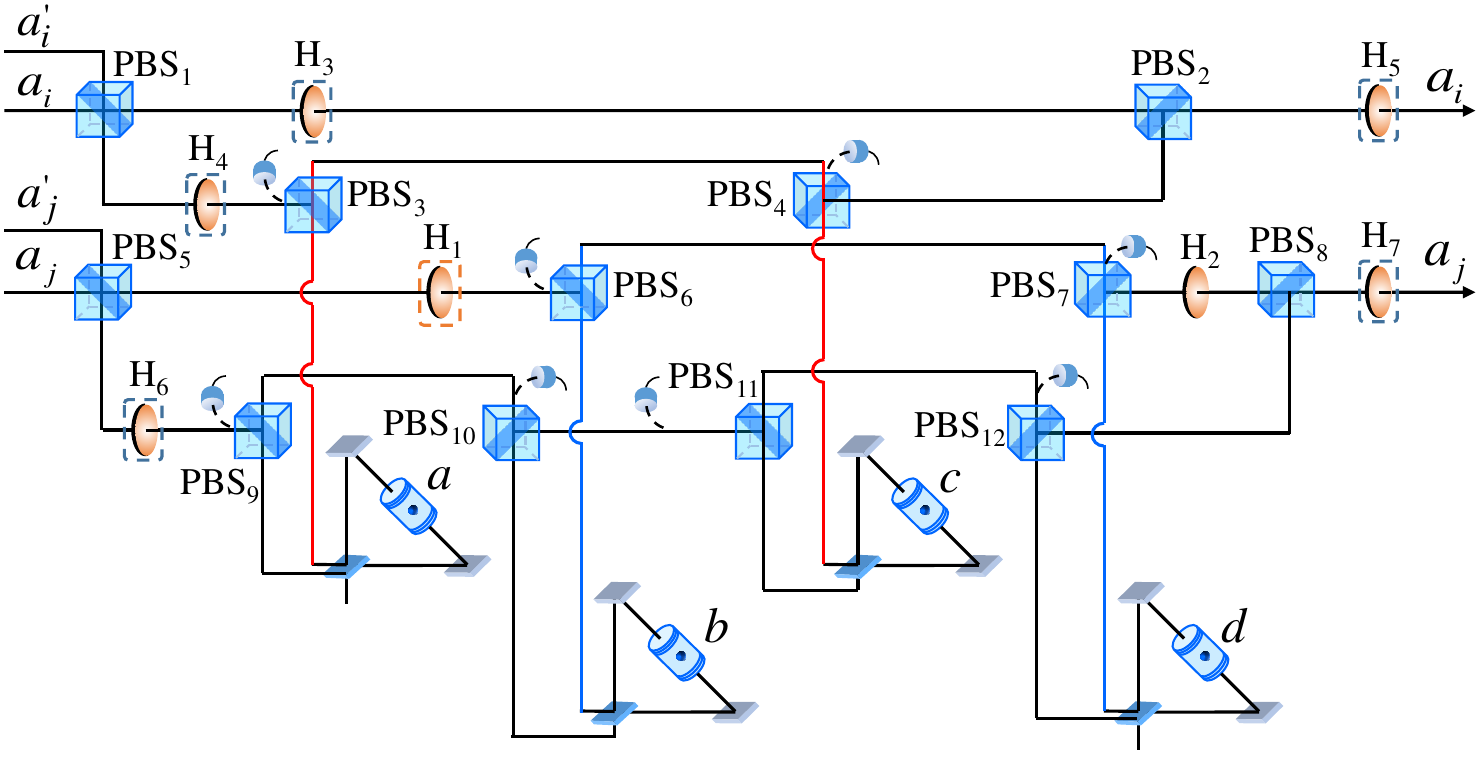}	\caption{Schematic diagram for  4D three-qudit  entanglement generation.} \label{fig4}	
		\end{center}
	\end{figure*}
	
	The $d$D 3-qudit maximally entangled GHZ state for hybrid system can be expressed in the form \cite{KerrGHZ}
	\begin{eqnarray}\label{eq24}
		|\varphi_{kpq}\rangle &=&\frac{1}{\sqrt{d}}\sum_{l=0}^{d-1}e^{i2\pi lk/d}|l\rangle\otimes|l\oplus p\rangle\otimes|l\oplus q\rangle_{123}.
	\end{eqnarray}
	Here, $k (k= 0, 1, 2, ..., d - 1)$ denotes the relative phase information, while $p$ and $q$  $(p, q = 0, 1, 2, ..., d - 1)$ denote the spin of emitters information of the $d$D entanglement. Besides, $l\oplus p\equiv(l\oplus p)\mathrm{mod}d$, $l\oplus q\equiv(l\oplus q)\mathrm{mod}d$. The subscripts 1, 2 and 3 denote the three entangled qudits.
	When $k=0, 1,2,3$,  the 4D three-qudit maximally entangled states can be written as
	\begin{eqnarray}\label{eq25}
		\left|\varphi_{0pq}\right\rangle &=& \frac12(|0\rangle \otimes|0\oplus p\rangle\otimes|0\oplus q\rangle\nonumber\\
		&&+|1\rangle \otimes|1\oplus p\rangle\otimes|1\oplus q\rangle\nonumber\\
		&&+|2\rangle \otimes|2\oplus p\rangle\otimes|2\oplus q\rangle\nonumber\\
		&&+|3\rangle \otimes|3\oplus p\rangle\otimes|3\oplus q\rangle)_{123}, \nonumber\\
		\left|\varphi_{1pq}\right\rangle &=& \frac12(|0\rangle \otimes|0\oplus p\rangle\otimes|0\oplus q\rangle\nonumber\\
		&&+e^{i\frac{\pi}{2}}|1\rangle \otimes|1\oplus p\rangle\otimes|1\oplus q\rangle\nonumber\\
		&&+e^{i\pi}|2\rangle \otimes|2\oplus p\rangle\otimes|2\oplus q\rangle\nonumber\\
		&&+e^{i\frac{3\pi}{2}}|3\rangle \otimes|3\oplus p\rangle\otimes|3\oplus q\rangle)_{123}, \nonumber\\
		\left|\varphi_{2pq}\right\rangle &=& \frac12(|0\rangle \otimes|0\oplus p\rangle\otimes|0\oplus q\rangle\nonumber\\
		&&+e^{i\pi}|1\rangle \otimes|1\oplus p\rangle\otimes|1\oplus q\rangle\nonumber\\
		&&+|2\rangle \otimes|2\oplus p\rangle\otimes|2\oplus q\rangle\nonumber\\
		&&+e^{i\pi}|3\rangle \otimes|3\oplus p\rangle\otimes|3\oplus q\rangle)_{123}, \nonumber\\
		\left|\varphi_{3pq}\right\rangle &=& \frac12(|0\rangle \otimes|0\oplus p\rangle\otimes|0\oplus q\rangle\nonumber\\
		&&+e^{i\frac{3\pi}{2}}|1\rangle \otimes|1\oplus q\rangle\otimes|1\oplus q\rangle\nonumber\\
		&&+e^{i\pi}|2\rangle \otimes|2\oplus p\rangle\otimes|2\oplus q\rangle\nonumber\\
		&&+e^{i\frac{\pi}{2}}|3\rangle \otimes|3\oplus p\rangle\otimes|3\oplus q\rangle)_{123}.
	\end{eqnarray}
Consistent with the generation proposal of the 4D 2-qudit entangled state,
	the generation of the 4D 3-qudit entangled state demands four emitters $a$, $b$, $c$, and $d$, with the schematic diagram shown in Fig. \ref{fig4}. Assuming the preparation process of the single photon is the same as that of the above two-qudit entangled state, shown in Fig. \ref{fig2}(a),
	four emitters $a$, $b$, $c$, and $d$ are in the initial states $|+\rangle_a$, $|+\rangle_b$, $|+\rangle_c$, and $|+\rangle_d$, respectively,
	where two emitters $c$ and $d$ are used to encode the third 4D qudit, i.e.,  $|++\rangle_{cd}=|0\rangle_{3}$, $|+-\rangle_{cd}=|1\rangle_{3}$,
	$|-+\rangle_{cd}=|2\rangle_{3}$, and $|--\rangle_{cd}=|3\rangle_{3}$.
	Similarly, when the three-qudit entangled state $|\varphi_{000}\rangle$ need to be implemented,
	two paths $a_{1}$ and $a_{2}$ should be connected to two input paths $a_{i} (i=1)$ and  $a_{j} (j=2)$, respectively.
	After that, the subsequent process will be performed: the $|H\rangle$-polarized photon in path $a_{1}$ is directly transmitted by PBS$_{1}$ and PBS$_{2}$ without interacting with the four emitters. The $|V\rangle$-polarized photon in  path $a_{1}$ undergoes a series of operations, i.e., PBS$_{1}$, PBS$_{3}$$\rightarrow$ interaction with emitter $a\rightarrow$ PBS$_{3}$,  PBS$_{4}\rightarrow$ interaction with emitter $c\rightarrow$ PBS$_{4}$, PBS$_{2}$ in sequence, and converges at the path $a_{1}$. On the other hand, the $|H\rangle$-polarized photon in path $a_{2}$ undergoes a series of operations, i.e., H$_{1}\rightarrow$ PBS$_{6}\rightarrow$ interaction with emitter $b\rightarrow$ PBS$_{6}$,  PBS$_{7}\rightarrow$ interaction with emitter $d\rightarrow$ PBS$_{7}\rightarrow$ H$_{2}\rightarrow $PBS$_{8}$ in turn, meanwhile the $|V\rangle$-polarized photon in path $a_{2}$ undergoes a series of operations, i.e., PBS$_{5}\rightarrow$ PBS$_{9}\rightarrow$ interaction with emitter $a\rightarrow$ PBS$_{9}$, PBS$_{10}\rightarrow$ interaction with emitter $b\rightarrow$   PBS$_{10}$, PBS$_{11}\rightarrow$  interaction with emitter $c \rightarrow$ PBS$_{11}$, PBS$_{12}\rightarrow$ interaction with emitter $d\rightarrow$ PBS$_{12}$, PBS$_{8}$  in turn, and at last two pulses converge into the path $a_{2}$. The composite state $\frac{1}{2}(|H\rangle+|V\rangle)\otimes(|a_{1}\rangle+|a_{2}\rangle) \otimes|+\rangle_a|+\rangle_b|+\rangle_c|+\rangle_d$  is transformed into
	\begin{eqnarray}\label{eq26}
		|\Psi\rangle_{1}&=&\frac{1}{2}(|Ha_1\rangle|+\rangle_a|+\rangle_b|+\rangle_c|+\rangle_d\nonumber\\
		&&+r^{2}|Ha_2\rangle|+\rangle_a|-\rangle_b|+\rangle_c|-\rangle_d\nonumber\\
		&&+r^{2}|Va_1\rangle|-\rangle_a|+\rangle_b|-\rangle_c|+\rangle_d\nonumber\\
		&&+r^{4}|Va_2\rangle|-\rangle_a|-\rangle_b|-\rangle_c|-\rangle_d)\nonumber\\
		&=&\frac{1}{2}(|000\rangle+r^{2}|111\rangle+r^{2}|222\rangle\nonumber\\
		&&+r^{4}|333\rangle)_{123}.
	\end{eqnarray}
	Under ideal scattering conditions $r = -1$, the state $|\Psi\rangle_{1}$ undergoes transformation to the 4D three-qudit maximally entangled state $\left|\varphi_{000}\right\rangle$.
	
\begin{figure}[htbp]
		\begin{center}	\includegraphics[width=8 cm,angle=0]{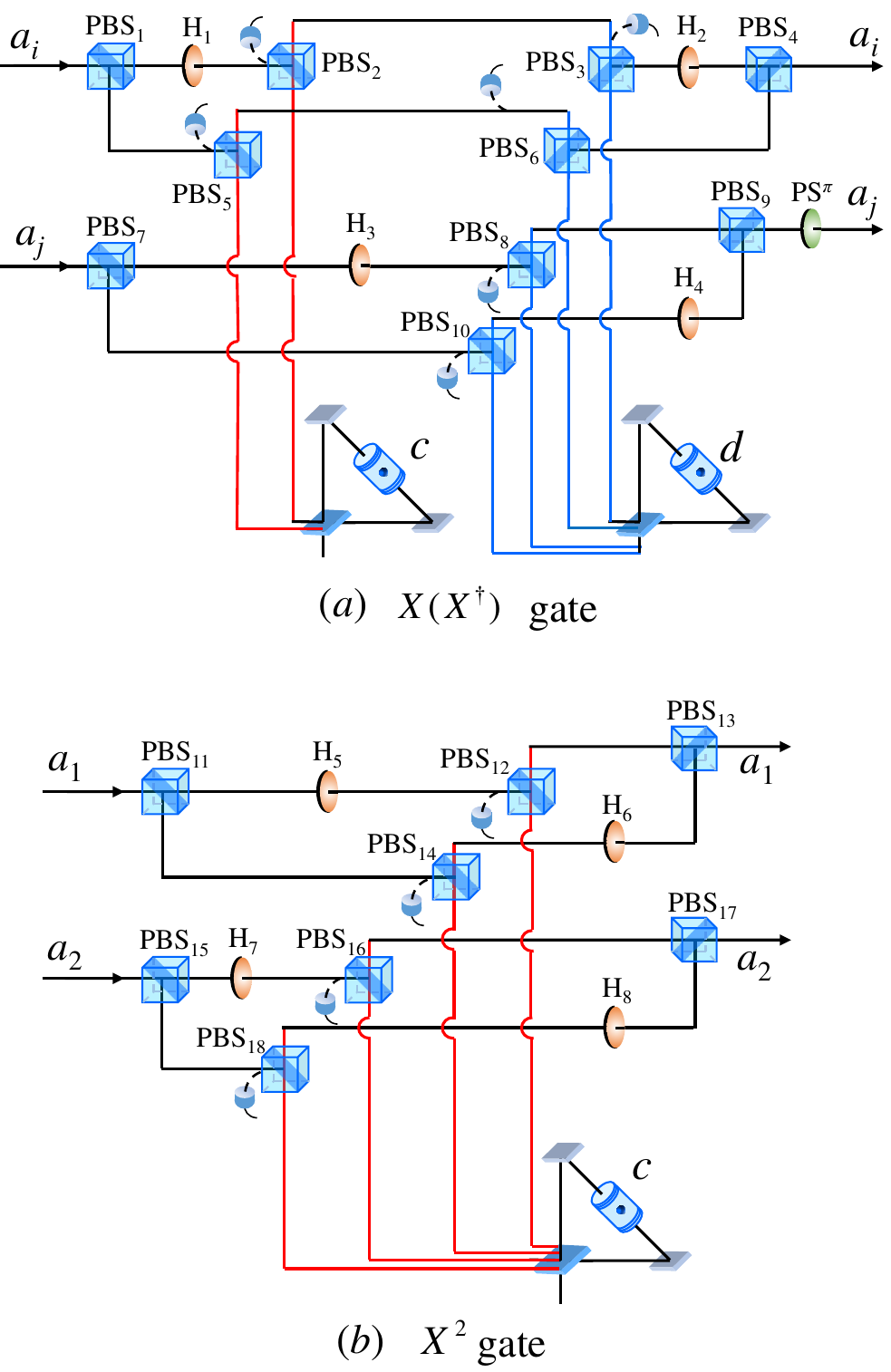}	\caption{(a)The $X$ or $X^{\dag}$ gate for the third qudit constructed by linear-optics elements and two emitters $c$ and $d$. For the 4D $X$ gate, two paths $a_{1}$ and $a_{2}$ should be connected to two input paths $a_{j} (j=1)$ and  $a_{i} (i=2)$, respectively. Otherwise, for the 4D $X^{\dag}$ gate, two paths $a_{1}$ and $a_{2}$ should be connected to two input paths $a_{i} (i=1)$ and  $a_{j} (j=2)$, respectively.
				(b)The 4D $X^{2}$ gate for the third qudit constructed by one emitter $c$. } \label{fig5}	
		\end{center}
	\end{figure}

	By controlling the different input ports of paths $a_{1}$ and $a_{2}$, and flexibly using half-wave plates, the same as four cases of proposal of 4D two-qudit entanglement generation, the 4D three-qudit  state $|\varphi_{000}\rangle$, $|\varphi_{011}\rangle$,
	$|\varphi_{022}\rangle$, and
	$|\varphi_{033}\rangle$ can be obtained sequentially. Further, we can convert the generated states $|\varphi\rangle_{000}$, $|\varphi_{011}\rangle$,
	$|\varphi_{022}\rangle$, and
	$|\varphi_{033}\rangle$  any desired 4D three-qudit maximally entangled state with $k=0$ in Eq. (\ref{eq25}), by the 4D single-qudit Pauli $X^{m}$  gate, expressed by	$X^{m}|l\rangle=|l\oplus m\rangle$ ($m,l= 0, 1, 2, 3$). Distinct Pauli $X^{m}$$(m=0, 1, 2, 3$)
	gates have been designed that perform cyclic operations, shifting each quantum state to its $m$-th nearest state along the clockwise orientation, namely, the $X^{0}$ ($I$) gate, $X^{1}$ ($X$) gate, $X^{2}$ gate, and $X^{3}$ ($X^{\dagger}$) gate, as shown in Fig. \ref{fig5}.
	The matrix representations of 4D Pauli $X$ gates with the specified $\{|0\rangle_{3}, |1\rangle_{3}, |2\rangle_{3}, |3\rangle_{3}\}$ bases, can also expressed as
	\begin{eqnarray}\label{eq27}
		I = \left( \begin{array}{cccc}
			1 & 0 & 0 & 0 \\
			0	& 1 & 0 &  0\\
			0	& 0 & 1 & 0 \\
			0	& 0 & 0 & 1 \\
		\end{array} \right),\;\; X = \left( \begin{array}{cccc}
			0	& 1 & 0 & 0 \\
			0	& 0 & 1&0  \\
			0	& 0 & 0 & 1 \\
			1	& 0 & 0 & 0 \\
		\end{array} \right),\nonumber\\
		X^{2} = \left( \begin{array}{cccc}
			0	&  0&1 & 0 \\
			0	& 0  & 0 & 1 \\
			1& 0 &0  &0  \\
			0	& 1 & 0 &  0\\
		\end{array} \right),\;\;
		X^{\dag} = \left( \begin{array}{cccc}
			0	& 0 &0  & 1 \\
			1&  0& 0 &0  \\
			0	& 1 &0  & 0 \\
			0	& 0 & 1 &0  \\
		\end{array} \right).
	\end{eqnarray}
	
	The schematic diagram of 4D $X$ gate is shown in Fig. \ref{fig5}(a), which corresponds to paths $a_{1}$ and $a_{2}$ connected with paths $a_{j}$($j=1$) and $a_{i}$($i=2$), respectively. On the contrary, it corresponds to 4D $X^{\dag}$ gate if two paths $a_{1}$ and $a_{2}$ connected with two paths $a_{i}$($i=1$)and $a_{j}$($j=2$), respectively. Supposed that the input entangled state of the photon is $|\varphi_{000}\rangle$. The operations performed on the $|H\rangle$-polarized photon in path $a_2$ are as follows: PBS$_{1}\rightarrow$ H$_{1}\rightarrow$ PBS$_{2}\rightarrow$ interaction with emitters $c$ and $d\rightarrow$ H$_{2}\rightarrow$ PBS$_{4}$. Similarly, the operations on the $|V\rangle$-polarized photon in path $a_2$ are:  PBS$_{1} \rightarrow$ PBS$_{5}\rightarrow$ interaction with emitters $c$ and $d\rightarrow$ PBS$_{4}$.
	For the $|H\rangle$-polarized photon in path $a_1$, the sequence of operations contain: PBS$_{7}\rightarrow$ H$_{3}\rightarrow$ PBS$_{8}\rightarrow$ interaction with emitter $d\rightarrow$ PBS$_{8}\rightarrow$ PBS$_{9}$.
	For the $|V\rangle$-polarized photon in path $a_1$, the operations are: PBS$_{7}\rightarrow$ PBS$_{10}\rightarrow$  interaction with emitter $d\rightarrow$ PBS$_{10}\rightarrow$ H$_{4}\rightarrow$ PBS$_{9}\rightarrow$PS$^{\pi}$. After that, the state $|\varphi_{000}\rangle$ is converted into
	\begin{eqnarray}\label{eq28}
		|\varphi_{000}\rangle
		&\xrightarrow{X}&\frac{1}{2}(-r|Ha_1\rangle|+\rangle_a|+\rangle_b|+\rangle_c|-\rangle_d\nonumber\\
		&&+r^{2}|Ha_2\rangle|+\rangle_a|-\rangle_b|-\rangle_c|+\rangle_d\nonumber\\
		&&-r|Va_1\rangle|-\rangle_a|+\rangle_b|-\rangle_c|-\rangle_d\nonumber\\
		&&+r^{2}|Va_2\rangle|-\rangle_a|-\rangle_b|+\rangle_c|+\rangle_d)\nonumber\\
		&=&\frac{1}{2}(-r|001\rangle+r^{2}|112\rangle-r|223\rangle\nonumber\\
		&&+r^{2}|330\rangle)_{12}.
	\end{eqnarray}
	Under a ideal scattering process, namely, $r = -1$, the above state
	is converted into the 4D three-qudit maximally entangled state $|\varphi_{001}\rangle$.
	Further, for 4D $X^{\dag}$ gate, shown in Fig. \ref{fig5}(a),  the  transformation is evolved into
	\begin{eqnarray}\label{eq29}
		\left|\varphi_{000}\right\rangle
		&\xrightarrow{X^{\dag}}&\frac{1}{2}(r^{2}|Ha_1\rangle|+\rangle_a|+\rangle_b|-\rangle_c|-\rangle_d\nonumber\\
		&&-r|Ha_2\rangle|+\rangle_a|-\rangle_b|+\rangle_c|+\rangle_d\nonumber\\
		&&+r^{2}|Va_1\rangle|-\rangle_a|+\rangle_b|+\rangle_c|-\rangle_d\nonumber\\
		&&-r|Va_2\rangle|-\rangle_a|-\rangle_c|-\rangle_a|+\rangle_d)\nonumber\\
		&=&\frac{1}{2}(r^{2}|003\rangle-r|110\rangle+r^{2}|221\rangle\nonumber\\
		&&-r|332\rangle)_{12}.
	\end{eqnarray}
	For a ideal scattering characterized by $r = -1$, the above state
	is converted into the state $|\varphi_{003}\rangle$.
	\begin{table*}[htbp]
		\centering
		\caption{The operations required for arbitrary 4D 3-qudit maximally entangled states.} \label{Table2}
		\begin{ruledtabular}
		\begin{tabular}{ccccccccccccccccc}
			\diagbox{$k$}{$pq$} & 00 & 01 & 02 & 03 & 10 & 11 & 12 & 13 & 20 & 21 & 22 & 23 & 30 & 31 & 32 & 33 \\
			\hline
			0 & $Z^{0}$ & $X$ & $X^{2}$ & $X^{\dagger}$ & $X^{\dagger}$ & $Z^{0}$ & $X$ & $X^{2}$ & $X^{2}$ & $X^{\dagger}$ & $Z^{0}$ & $X$ & $X$ & $X^{2}$ & $X^{\dagger}$ & $Z^{0}$ \\
			1 & $Z$ & $XZ$ & $X^{2}Z$ & $X^{\dagger}Z$ & $X^{\dagger}Z$ & $Z$ & $XZ$ & $X^{2}Z$ & $X^{2}Z$ & $X^{\dagger}Z$ & $Z$ & $XZ$ & $XZ$ & $X^{2}Z$ & $X^{\dagger}Z$ & $Z$ \\
			2 & $Z^{2}$ & $XZ^{2}$ & $X^{2}Z^{2}$ & $X^{\dagger}Z^{2}$ & $X^{\dagger}Z^{2}$ & $Z^{2}$ & $XZ^{2}$ & $X^{2}Z^{2}$ & $X^{2}Z^{2}$ & $X^{\dagger}Z^{2}$ & $Z^{2}$ & $XZ^{2}$ & $XZ^{2}$ & $X^{2}Z^{2}$ & $X^{\dagger}Z^{2}$ & $Z^{2}$ \\
			3 & $Z^{\dagger}$ & $XZ^{\dagger}$ & $X^{2}Z^{\dagger}$ & $X^{\dagger}Z^{\dagger}$ & $X^{\dagger}Z^{\dagger}$ & $Z^{\dagger}$ & $XZ^{\dagger}$ & $X^{2}Z^{\dagger}$ & $X^{2}Z^{\dagger}$ & $X^{\dagger}Z^{\dagger}$ & $Z^{\dagger}$ & $XZ^{\dagger}$ & $XZ^{\dagger}$ & $X^{2}Z^{\dagger}$ & $X^{\dagger}Z^{\dagger}$ & $Z^{\dagger}$ \\
\end{tabular}
\end{ruledtabular}
	\end{table*}
	
	Fig. \ref{fig5}(b) shows the schematic diagram of the Pauli 4D $X^{2}$ gate. The $|H\rangle$-polarized photon in path $a_1$ undergoes the following sequence of operations: PBS$_{11}\rightarrow$ H$_{5}\rightarrow$ PBS$_{12}\rightarrow$ interaction with emitter $c\rightarrow$ PBS$_{12}\rightarrow$ PBS$_{13}$.
	The $|V\rangle$-polarized photon in path $a_1$ undergoes: PBS$_{11}\rightarrow$ PBS$_{14}\rightarrow$ interaction with emitter $c \rightarrow$ PBS$_{14}\rightarrow$ H$_{6}\rightarrow$ PBS$_{13}$.
	For the $|H\rangle$-polarized photon in path $a_2$, the operations are: PBS$_{15}\rightarrow$ H$_{7}\rightarrow$  PBS$_{16}\rightarrow$ interaction with emitter $a\rightarrow$ PBS$_{16}\rightarrow$ PBS$_{17}$.
	The $|V\rangle$-polarized photon in path $a_2$ undergoes:  PBS$_{15}\rightarrow$ PBS$_{18}\rightarrow$ interaction with emitter $c\rightarrow$ PBS$_{18}\rightarrow$ H$_{8}\rightarrow$ PBS$_{17}$. After that, the state $|\varphi_{000}\rangle$ is converted into
	\begin{eqnarray}\label{eq30}
		\left|\varphi_{000}\right\rangle
		&\xrightarrow{X^{2}}&\frac{1}{2}(r|Ha_1\rangle|+\rangle_a|+\rangle_b|-\rangle_c|+\rangle_d\nonumber\\
		&&+r|Ha_2\rangle|+\rangle_a|-\rangle_b|-\rangle_c|-\rangle_d\nonumber\\
		&&+r|Va_1\rangle|-\rangle_a|+\rangle_b|+\rangle_c|+\rangle_d\nonumber\\
		&&+r|Va_2\rangle|-\rangle_a|-\rangle_c|+\rangle_a|-\rangle_d)\nonumber\\
		&=&\frac12 r(|002\rangle+|113\rangle+|220\rangle+|331\rangle)_{12}\nonumber\\
		&=&r|\varphi_{002}\rangle.
	\end{eqnarray}
	Obviously, the states $|\varphi_{012}\rangle$, $|\varphi_{013}\rangle$, and $|\varphi_{010}\rangle$ can be obtained from the state $|\varphi_{011}\rangle$ by applying the $X$ gate, $X^{2}$ gate, and $X^{\dagger}$ gate, respectively. Similarly, the states $|\varphi_{023}\rangle$, $|\varphi_{020}\rangle$, and $|\varphi_{021}\rangle$ can be obtained from the state $|\varphi_{022}\rangle$, and the states $|\varphi_{030}\rangle$, $|\varphi_{031}\rangle$, and $|\varphi_{032}\rangle$ can be obtained from the state $|\varphi_{033}\rangle$.
	By combining single-qudit Pauli $X^{m}$ and $Z^{m}$ gates, the arbitrary 4D three-qudit maximally entangled states in Eq. (\ref{eq15}) for hybrid system can be realized in the HD state space.
	Table \ref{Table3} presents the various 4D three-qudit maximally entangled states for hybrid system obtained by using Pauli $X^{m}$ and Pauli $Z^{m}$ gates.

	\section{4D $n$-Qudit entanglement generation for hybrid system}\label{sec4}
	\begin{figure*}[htbp]
		\centering
		\subfigure[]
		{
		\begin{minipage}{1\linewidth}
				\centering
				\includegraphics[width=0.65\linewidth]{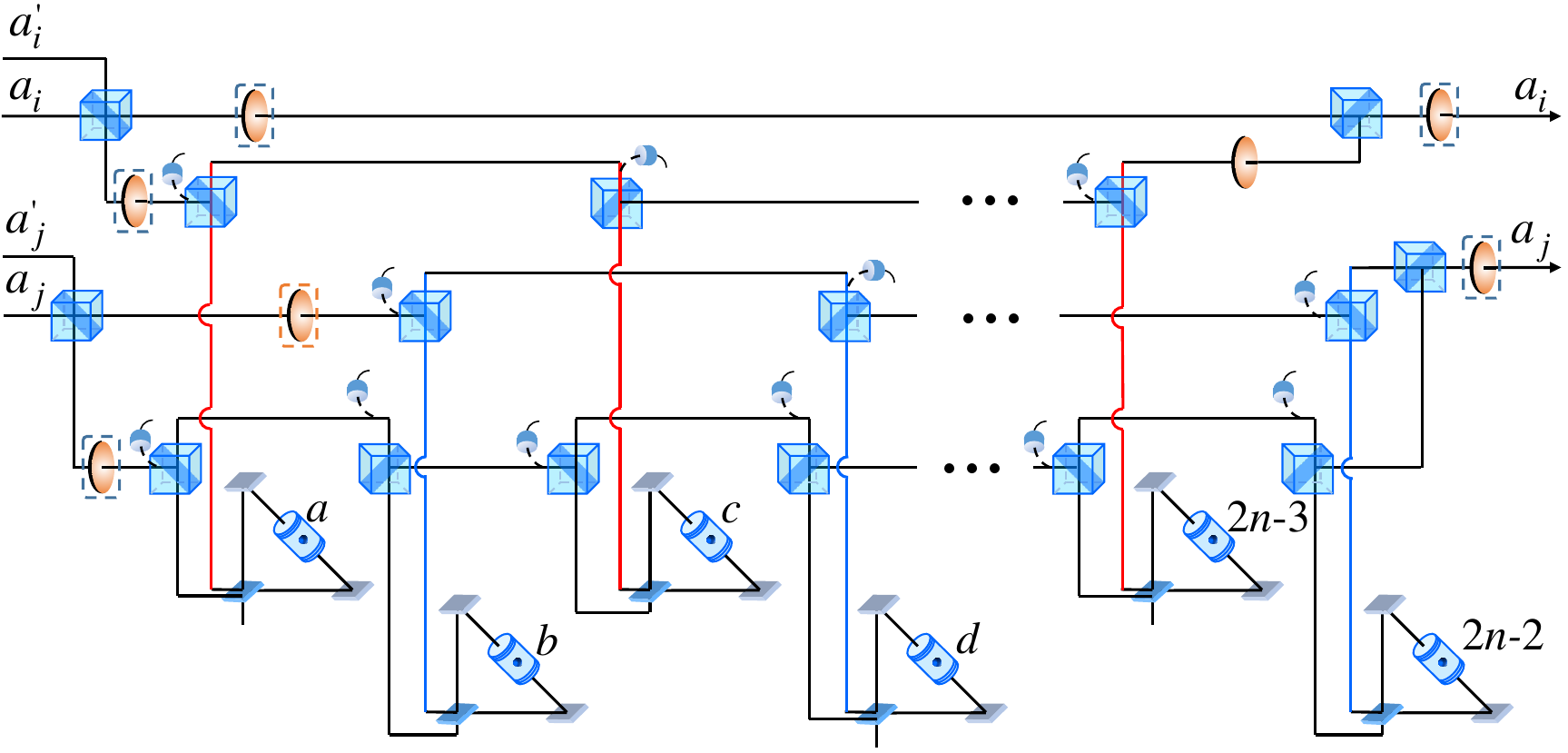}
				\label{fig6a}
		\end{minipage}}	
		
		\subfigure[]
		{
			\begin{minipage}{1\linewidth}
				\centering
				\includegraphics[width=0.65\linewidth]{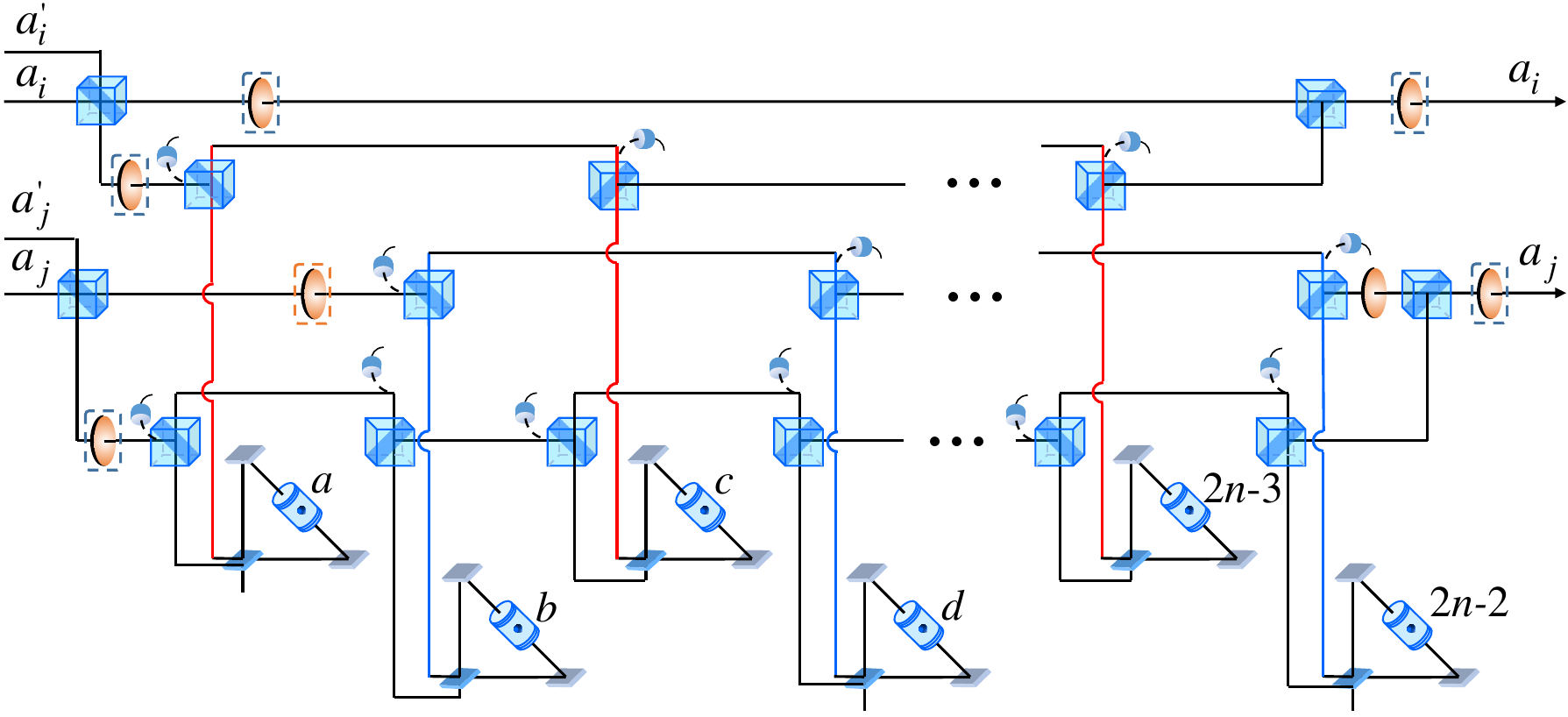}
				\label{fig6b}
		\end{minipage}}
		\caption{Schematic diagrams for generating 4D $n$-qudit maximally entangled states. (a)$n$ is odd, (b)$n$ is even.} \label{fig6}
	\end{figure*}
	As discussed in previous sections, we have investigated the generation processes of 4D two-qudit and 4D three-qudit entangled states.  Based on the aforementioned schemes, these can be generalized to $n$-qudit systems. Consequently,  we will demonstrate how to generate 4D $n$-qudit($n > 3$) entangled states for $n$-qudit system, with $2n-2$ waveguide-emitters systems illustrated in Fig. \ref{fig6}(a-b), where Fig. \ref{fig6}(a) corresponds to the case that $n$ is odd, while the scenario that $n$ is even in Fig. \ref{fig6}(b). For the hybrid system, the first 4D qudit is encoded on the hybrid polarization-path states of the single photon, while the rest 4D qudits are represented by two stationary emitters coupled to respective 1D waveguide.
	
	Assuming the preparation process of the single photon is the same as that of the above $2$-qudit entangled state and $3$-qudit entangled state,
	$2n-2$ emitters $a$, $b$, $c$, and $2n-2$ are in the initial states $|+\rangle_a$, $|+\rangle_b$, ... $|+\rangle_{2n-3}$,  and $|+\rangle_{2n-2}$, respectively.
	Similarly, for generating the $n$-qudit entangled state $|\varphi_{00...0}\rangle$, two paths $a_{1}$ and $a_{2}$ should be connected to two input paths $a_{i} (i=1)$ and  $a_{j} (j=2)$, respectively. Subsequently, the following series of operations are  performed:
	if the photon is in the state $|Ha_1\rangle$, it does not interact with $2n-2$ emitters; if the photon is in the state $|Ha_2\rangle$, it interacts with emitters $b$, $d$, ..., $2n-2$; if the photon is in the state $|Va_1\rangle$, it interacts with emitters $a$, $c$, ..., and $2n-3$; if the photon is in the state $|Va_2\rangle$, it interacts with emitters $a$, $b$, $c$, $d$, ..., $2n-3$, and $2n-2$. When $n$ is odd, the initial state $\frac{1}{2}(|H\rangle+|V\rangle)\otimes(|a_{1}\rangle+|a_{2}\rangle)\otimes |+\rangle_a |+\rangle_b ... |+\rangle_{2n-2}$ is changed into
	\begin{eqnarray}\label{eq31}
		|{{\Xi }_{1}}\rangle&=& \frac{1}{2}(|Ha_1\rangle|+\rangle_1|+\rangle_2...|+\rangle_{2n-3}|+\rangle_{2n-2}\nonumber\\
		&&+r^{n-1}|Ha_2\rangle|+\rangle_1|-\rangle_2...|+\rangle_{2n-3}|-\rangle_{2n-2}\nonumber\\
		&&+r^{n-1}|Va_1\rangle|-\rangle_1|+\rangle_2...|-\rangle_{2n-3}|+\rangle_{2n-2}\nonumber\\
		&&+r^{2n-2}|Va_2\rangle|-\rangle_1|-\rangle_2...|-\rangle_{2n-3}|-\rangle_{2n-2})\nonumber\\
		&=&\frac{1}{2}(|0\rangle^{\otimes n}+r^{n-1}|1\rangle^{\otimes n}+r^{n-1}|2\rangle^{\otimes n}\nonumber\\
		&&+r^{2n-2}|3\rangle^{\otimes n}).
	\end{eqnarray}
	When $n$ is even,  it is changed into
	\begin{eqnarray}\label{eq32}
		|{{\Xi }_{2}}\rangle&=& \frac{1}{2}(|Ha_1\rangle|+\rangle_1|+\rangle_2...|+\rangle_{2n-3}|+\rangle_{2n-2}\nonumber\\
		&&-r^{n-1}|Ha_2\rangle|+\rangle_1|-\rangle_2...|+\rangle_{2n-3}|-\rangle_{2n-2}\nonumber\\
		&&-r^{n-1}|Va_1\rangle|-\rangle_1|+\rangle_2...|-\rangle_{2n-3}|+\rangle_{2n-2}\nonumber\\
		&&+r^{2n-2}|Va_2\rangle|-\rangle_1|-\rangle_2...|-\rangle_{2n-3}|-\rangle_{2n-2})\nonumber\\
		&=&\frac{1}{2}(|0\rangle^{\otimes n}-r^{n-1}|1\rangle^{\otimes n}-r^{n-1}|2\rangle^{\otimes n}\nonumber\\
		&&+r^{2n-2}|3\rangle^{\otimes n}).
	\end{eqnarray}
	In the case of ideal scattering with $r = -1$,  the above state $|{{\Xi }_{1}}\rangle$ or $|{\Xi _{2}}\rangle$
	is converted into the 4D $n$-qudit maximally entangled states $\left|\varphi_{00...0}\right\rangle$ by the error-detected mechanism. By controlling the different input paths and flexibly using half-wave plates $45^{\circ}$, the $n$-qudit entangled state $|\varphi_{01\cdots1}\rangle$, $|\varphi_{02\cdots2}\rangle$, and $|\varphi_{03\cdots3}\rangle$ can be obtained. Furthermore, the 4D $n$-qudit entangled states $|\varphi_{00\cdots0}\rangle$, $|\varphi_{01\cdots1}\rangle$, $|\varphi_{02\cdots2}\rangle$, and $|\varphi_{03\cdots3}\rangle$ can be converted  into the rest forms of the 4D $n$-qudit maximal entangled state by  utilizing the 4D single-qudit $Z^{m} (m=1,2,3)$ gate for the first qudit and $X^{m}$ gate for the other qudits (except the second qudit).

	\section{8D two-qudit entanglement generation for hybrid system}\label{sec5}

	\begin{figure*}[htbp]
		\begin{center}	\includegraphics[width=14 cm,angle=0]{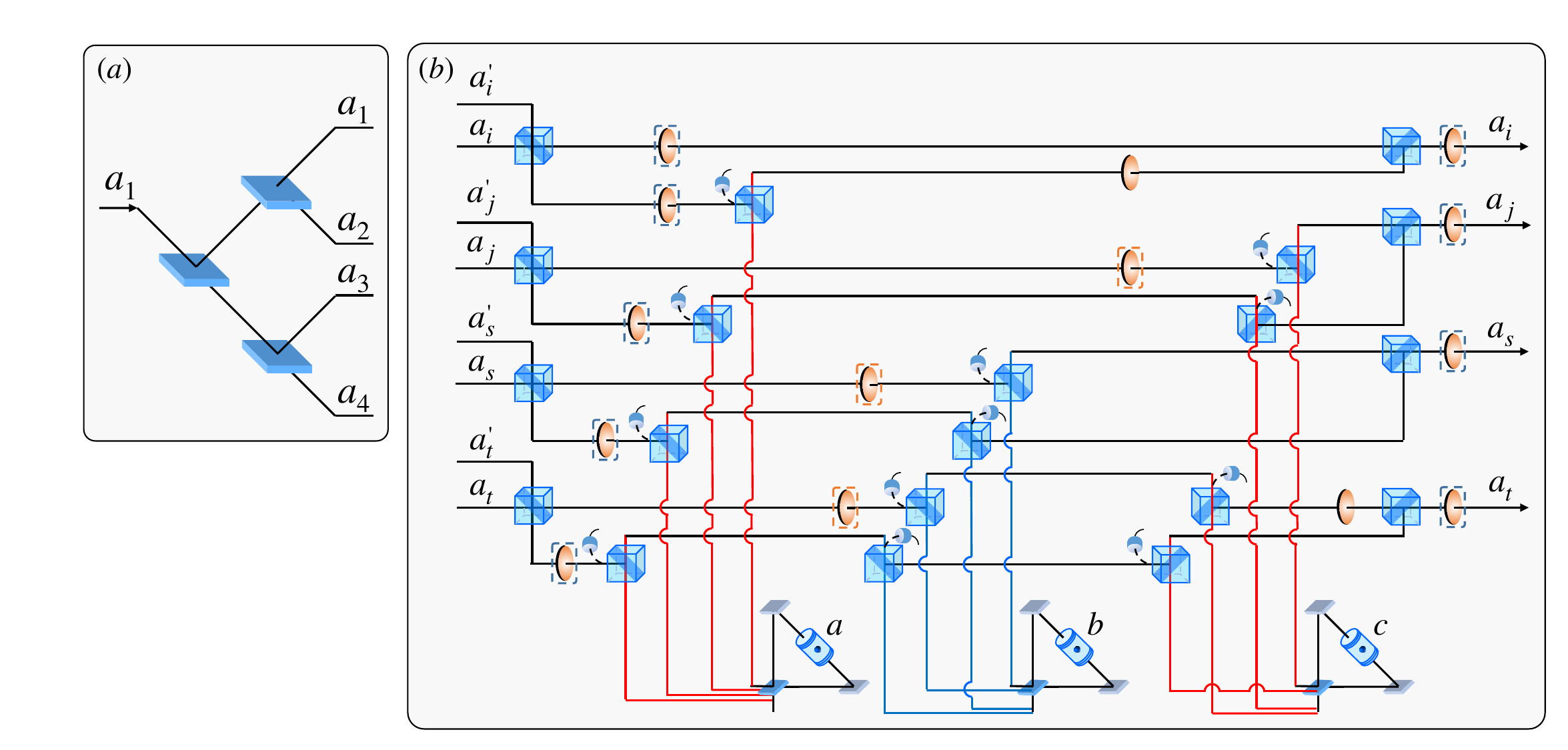}	\caption{(a)Schematic diagram for preparation of the initial state of the first qudit. (b)Schematic diagram for generating 8D two-qudit maximally entangled state.} \label{fig7}	
		\end{center}
	\end{figure*}
	So far, we have set up deterministic schemes for generating 4D 2-qudit entangled states, 3-qudit GHZ states, and $n$-qudit GHZ states in Sec. \ref{sec3} and Sec. \ref{sec4}. We discussed only the generation of 4D entangled states in the previous schemes, but this approach can also be extended to $d$D. In principle, the $d$D two-qudit maximally entangled state for hybrid system can be obtained by using $1+d/4$ emitters in 1D waveguide. Here, $\exists k \in \mathbb{Z}$, such that $d = 4k$.
	
	Taking $d = 8$ as an example, the two-qudit maximally entangled state, (where each qudit has 8D and thus a large Hilbert space) can be represented as:
	\begin{eqnarray}\label{eq33}
		|\phi_{pq}\rangle=\frac1{2\sqrt{2}}\sum_{l=0}^{7}e^{i\pi lp/4}|l\rangle\otimes|l\oplus q\rangle_{12}.
	\end{eqnarray}
	Specifically, the qudit 1 is mapped to the hybrid polarization-path states of the single photon, that is,
	\begin{eqnarray}\label{eq34}
		|Ha_{i}\rangle_{A} = |i-1\rangle_{1}, \;\; i \in \{1, 2, 3, 4\}, \nonumber \\
		|Va_{j}\rangle_{A} = |j+3\rangle_{1}, \;\; j \in \{1, 2, 3, 4\}.
	\end{eqnarray}
	And qudit 2 is mapped to the states of three emitters $a$, $b$, and $c$, that is
	\begin{eqnarray}\label{eq35}
		|+\rangle_{1}|+\rangle_{2}|+\rangle_{3}=|0\rangle_{2},\;\;|+\rangle_{1}|+\rangle_{2}|-\rangle_{3}=|1\rangle_{2},\nonumber\\
		|+\rangle_{1}|-\rangle_{2}|+\rangle_{3}=|2\rangle_{2},\;\;|+\rangle_{1}|-\rangle_{2}|-\rangle_{3}=|3\rangle_{2},\nonumber\\
		|-\rangle_{1}|+\rangle_{2}|+\rangle_{3}=|4\rangle_{2},\;\;|-\rangle_{1}|+\rangle_{2}|-\rangle_{3}=|5\rangle_{2},\nonumber\\
		|-\rangle_{1}|-\rangle_{2}|+\rangle_{3}=|6\rangle_{2},\;\;|-\rangle_{1}|-\rangle_{2}|-\rangle_{3}=|7\rangle_{2}.
	\end{eqnarray}
	The photon is divided into four paths $a_{1}$, $a_{2}$, $a_{3}$, and $a_{4}$ by three BSs as shown in Fig. \ref{fig7}(a). The initial state of the photon, $|{{X }_{0}}\rangle=\frac{1}{\sqrt{2}}(|H\rangle+|V\rangle)$, which is in the maximally superposed polarized state, is converted into $|{{X }_{1}}\rangle= \frac{1}{2\sqrt{2}}(|H\rangle+|V\rangle)\otimes(|a_{1}\rangle+|a_{2}\rangle+|a_{3}\rangle+|a_{4}\rangle)$.
Supposed that the states of three emitters $a$, $b$, and $c$ are $|+\rangle_1$, $|+\rangle_2$, and $|+\rangle_3$, respectively. As shown in Fig. \ref{fig7}(b), for achieving the 8D two-qudit entangled state $|\phi_{00}\rangle$, the four paths $a_{1}$, $a_{2}$, $a_{3}$, and $a_{4}$ should be connected to the paths $a_{i}$($i=1$), $a_{j}$($j=2$), $a_{s}$($s=3$), and $a_{t}$($t=4$), respectively.
If the photon is in the  state $|Ha_1\rangle$, it does not interact with three emitters. If it is in the state $|Ha_2\rangle$, it interacts with emitter $c$, while in the state $|Ha_3\rangle$, it interacts with emitter $b$, and in the state $|Ha_4\rangle$, it interacts with two emitters  $b$ and $c$. On the other hand, if the photon is in the state $|Va_1\rangle$, it interacts only with emitter $a$. For the state $|Va_2\rangle$, the interaction involves emitters $a$ and $c$; for the state $|Va_3\rangle$, it interacts with two emitters  $a$ and $b$; and for the state $|Va_4\rangle$, the photon interacts with all three emitters $a$, $b$, and $c$. Then, the state
$|{X _{1}}\rangle\otimes|+\rangle_1|+\rangle_2|+\rangle_3$ is transformed into
\begin{eqnarray}\label{eq36}
	|{X _{2}}\rangle &=& \frac{1}{2\sqrt{2}}(|00\rangle-r|11\rangle-r|22\rangle+r^{2}|33\rangle\nonumber\\
	&&
	-r|44\rangle+r^{2}|55\rangle+r^{2}|66\rangle-r^{3}|77\rangle)_{12}.
\end{eqnarray}
Under ideal scattering conditions, namely, $r = -1$, the above state $|{X _{2}}\rangle$
is converted into the 8D 2-qudit maximally entangled state $|\phi_{00}\rangle$ in Eq. (\ref{eq33}).
By controlling the different input paths and flexibly using wave plates $45^{\circ}$, we can obtain random 8D 2-qudit states, as shown in Tab. \ref{Table3}.
\begin{table}
	\centering
	\caption{The relationship between the all 8D 2-qudit entangled states and input ports of paths $a_{1}$ and $a_{2}$.} \label{Table3}
\begin{ruledtabular}
	\begin{tabular} {ccccccccc}
		& 	$\left|\varphi_{00}\right\rangle$ &	$\left|\varphi_{01}\right\rangle$ & 	$\left|\varphi_{02}\right\rangle$ &	$\left|\varphi_{03}\right\rangle$ & $\left|\varphi_{04}\right\rangle$& $\left|\varphi_{05}\right\rangle$& $\left|\varphi_{06}\right\rangle$&  $\left|\varphi_{07}\right\rangle$ \\
		\hline
		$a_{1}$ & $a_{i}$ & $a_{j}$ &$a_{s}$ & $a_{t}$ & $a'_{i}$ &$a'_{j}$ & $a'_{s}$&$a'_{t}$\\
		$a_{2}$ & $a_{j}$ & $a_{s}$ &$a_{t}$ &$a'_{i}$& $a'_{j}$& $a'_{s}$ &$a'_{t}$ &$a_{i}$\\
		$a_{3}$ & $a_{s}$ & $a_{t}$ &$a'_{i}$ & $a'_{j}$ & $a'_{s}$&$a'_{t}$ &$a_{i}$ &$a_{j}$\\
		$a_{4}$ & $a_{t}$ & $a'_{i}$ &$a'_{j}$ &$a'_{s}$ & $a'_{t}$ &$a_{i}$  & $a_{j}$ &$a_{s}$\\
\end{tabular}
\end{ruledtabular}
\end{table}

\section{Fidelity and Efficiency}\label{sec6}

\begin{figure}[htbp]
	\centering
	\subfigure[]
	{
		\begin{minipage}{0.7\linewidth}
			\centering
			\includegraphics[width=0.9\linewidth]{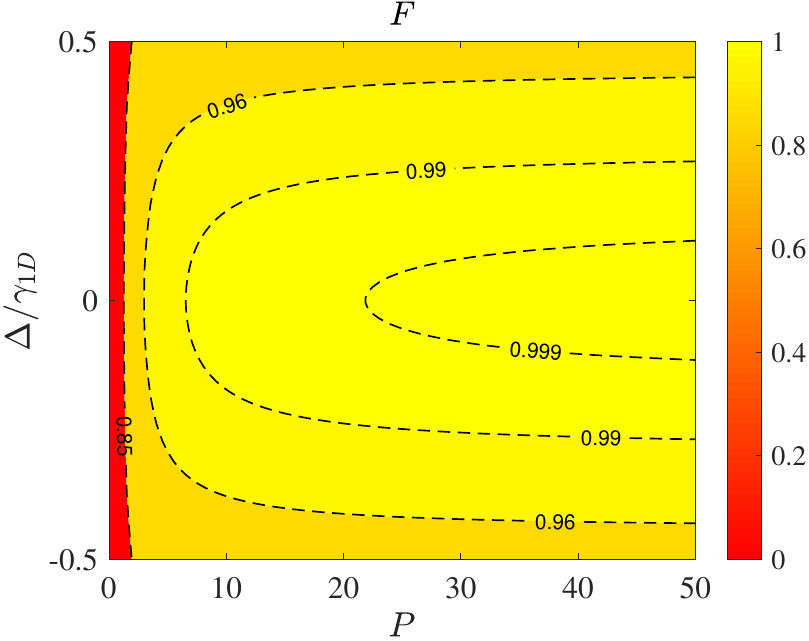}
			\label{fig8a}
	\end{minipage}}	
	\subfigure[]
	{
		\begin{minipage}{0.7\linewidth}
			\centering
			\includegraphics[width=0.9\linewidth]{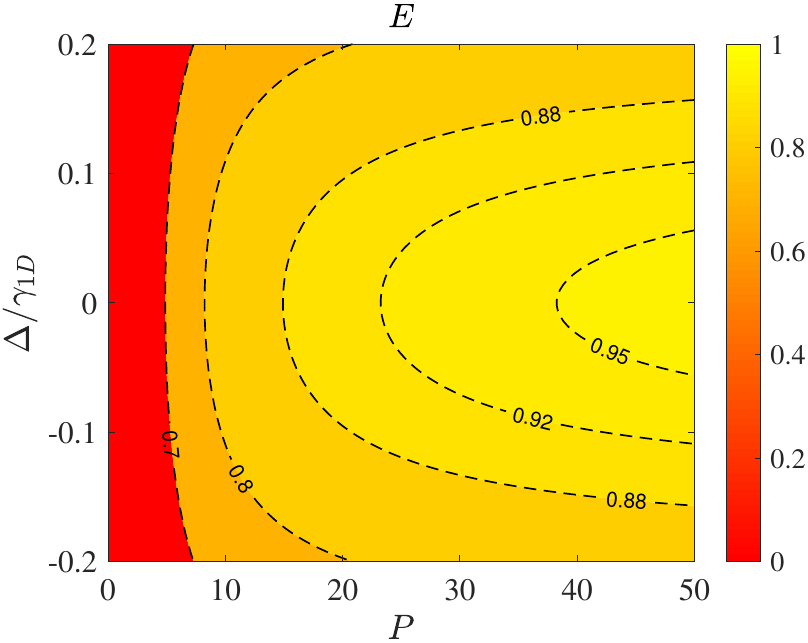}
			\label{fig8b}
	\end{minipage}}
	\caption{(a) The  fidelity $F$ and (b) the efficiency $E$ of the 4D two-qudit entanglement for hybrid system vs the Purcell factors $P$ and the frequency detuning $\Delta/{{\gamma }_{1D}}$.} \label{fig8}
\end{figure}
\begin{figure}[htbp]
	\centering
	\subfigure[]
	{
		\begin{minipage}{0.7\linewidth}
			\centering
			\includegraphics[width=0.9\linewidth]{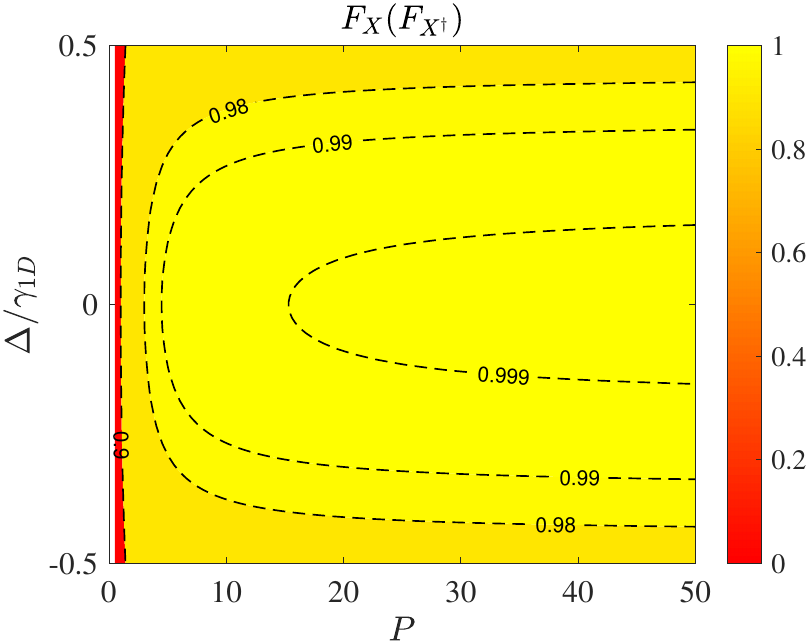}
			\label{fig9a}
	\end{minipage}}	
	\subfigure[]
	{
		\begin{minipage}{0.7\linewidth}
			\centering
			\includegraphics[width=0.9\linewidth]{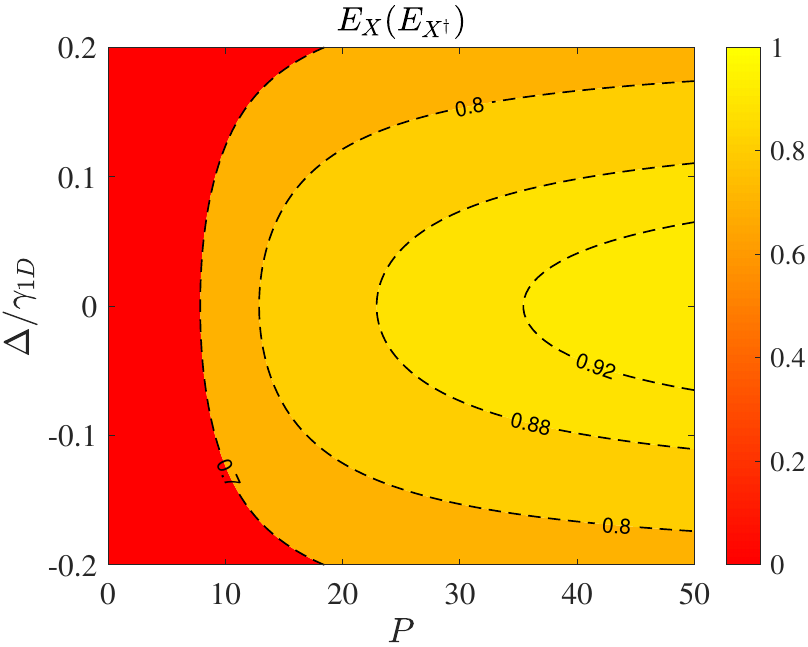}
			\label{fig9b}
	\end{minipage}}
	\subfigure[]
{
	\begin{minipage}{0.7\linewidth}
		\centering
		\includegraphics[width=0.9\linewidth]{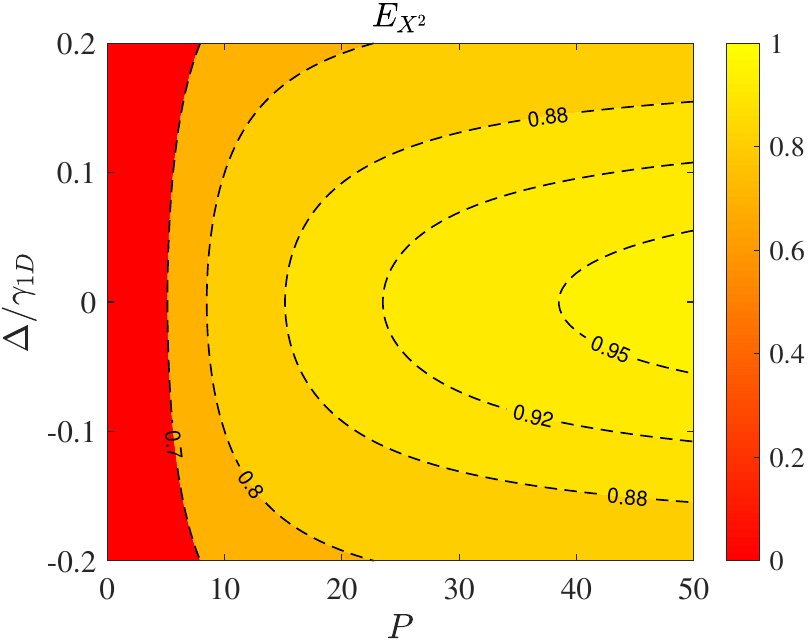}
		\label{fig9c}
\end{minipage}}
	\caption{(a) The  fidelity $F_{X}$($F_{X^{\dagger}}$) and (b) the efficiency $E_{X}$($E_{X^{\dagger}}$) and (c) the efficiency $E_{X^{2}}$ of the single-qudit $X^{m}$ gate for hybrid system vs the Purcell factors $P$ and the frequency detuning $\Delta/{{\gamma }_{1D}}$.} \label{fig9}
\end{figure}
\begin{figure}[htbp]
	\centering
	\subfigure[]
	{
		\begin{minipage}{0.7\linewidth}
			\centering
			\includegraphics[width=0.9\linewidth]{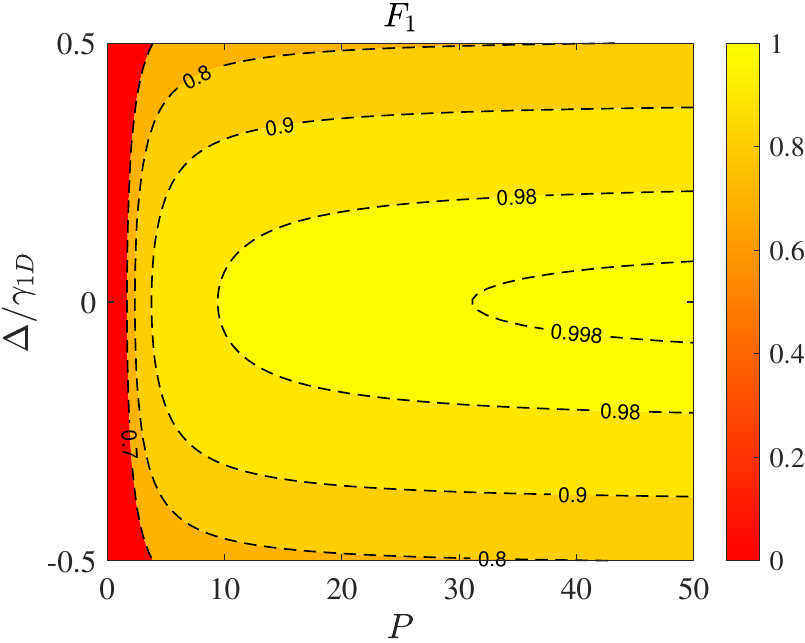}
			\label{fig10a}
	\end{minipage}}	
	\subfigure[]
	{
		\begin{minipage}{0.7\linewidth}
			\centering
			\includegraphics[width=0.9\linewidth]{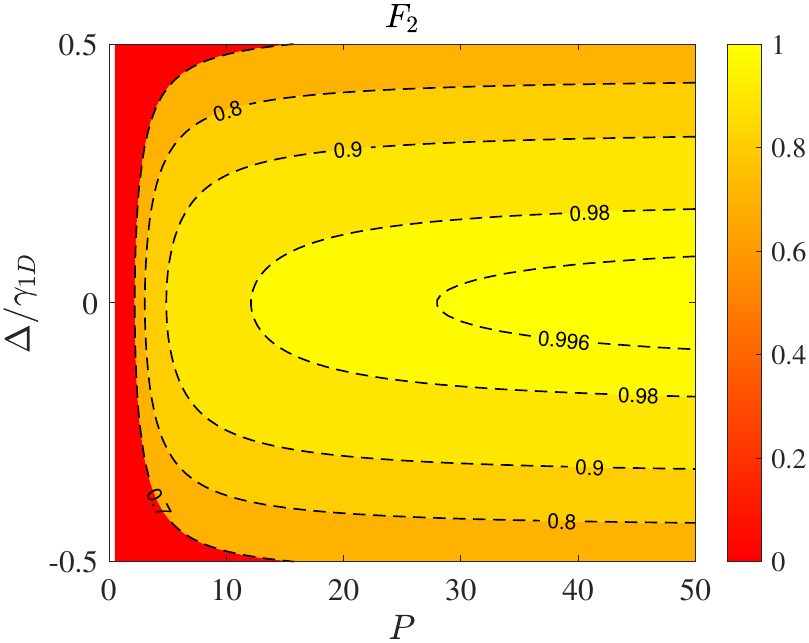}
			\label{fig10b}
	\end{minipage}}
	\subfigure[]
	{
		\begin{minipage}{0.7\linewidth}
			\centering
			\includegraphics[width=0.9\linewidth]{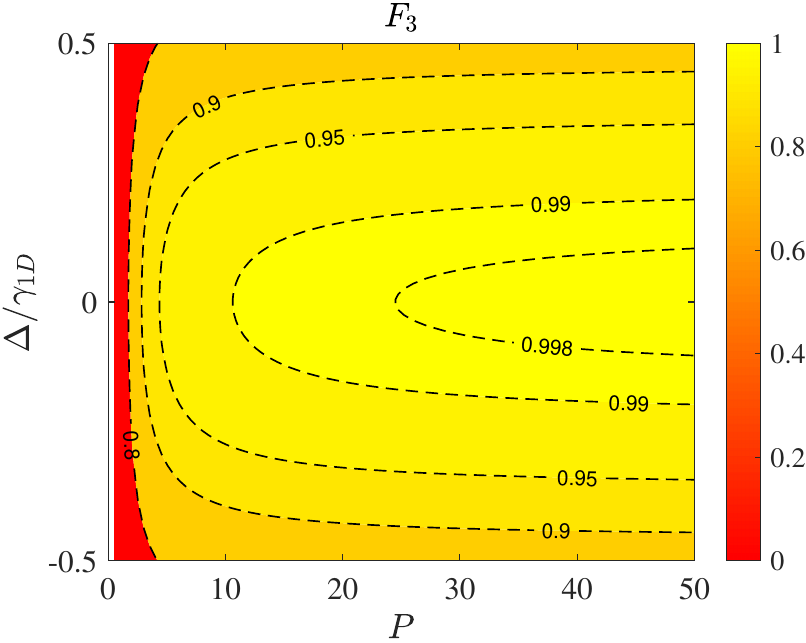}
			\label{fig10c}
	\end{minipage}}
	\caption{(a) The fidelity $F_{1}$ of the three-qudit entangled state  $|\varphi_{000}\rangle$ (or $|\varphi_{011}\rangle$, $|\varphi_{022}\rangle$, $|\varphi_{033}\rangle$, $|\varphi_{002}\rangle$, $|\varphi_{013}\rangle$, $|\varphi_{020}\rangle$,  $|\varphi_{031}\rangle$); (b) the fidelity $F_{2}$ of the state  $|\varphi_{001}\rangle$ (or $|\varphi_{010}\rangle$, $|\varphi_{023}\rangle$,  $|\varphi_{032}\rangle$); (c) the fidelity $F_{3}$ of the state $|\varphi_{003}\rangle$ (or $|\varphi_{012}\rangle$, $|\varphi_{021}\rangle$, $|\varphi_{030}\rangle$) vs the Purcell factors $P$ and the frequency detuning $\Delta/{{\gamma }_{1D}}$.} \label{fig10}
\end{figure}
\begin{figure}[htbp]
	\centering
	\subfigure[]
	{
		\begin{minipage}{0.75\linewidth}
			\centering
			\includegraphics[width=0.9\linewidth]{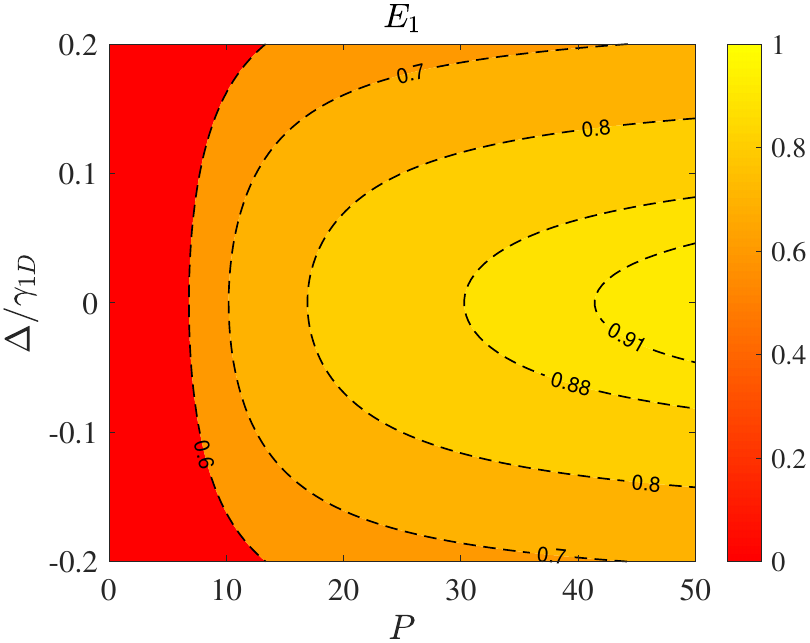}
			\label{fig11a}
	\end{minipage}}	
	\subfigure[]
	{
		\begin{minipage}{0.75\linewidth}
			\centering
			\includegraphics[width=0.9\linewidth]{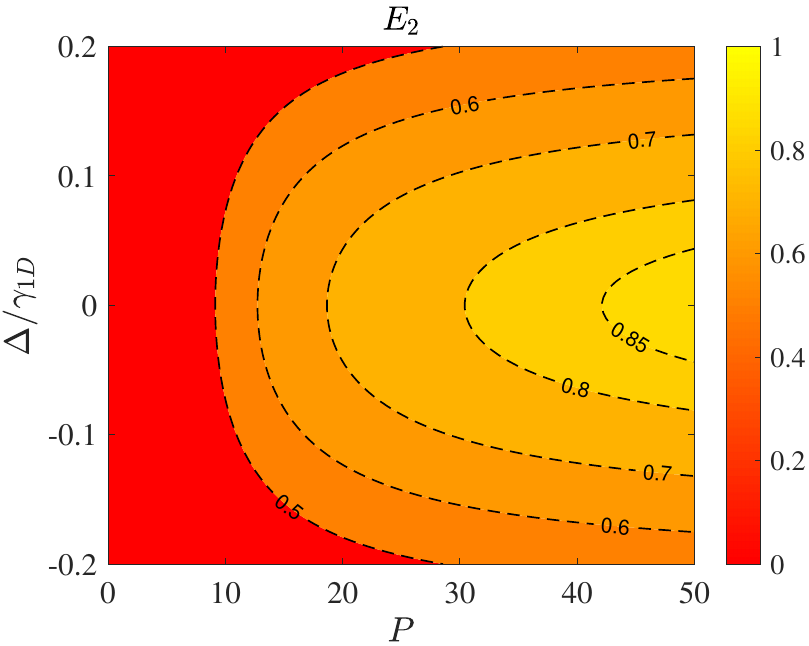}
			\label{fig11b}
	\end{minipage}}
	\subfigure[]
	{
		\begin{minipage}{0.75\linewidth}
			\centering
			\includegraphics[width=0.9\linewidth]{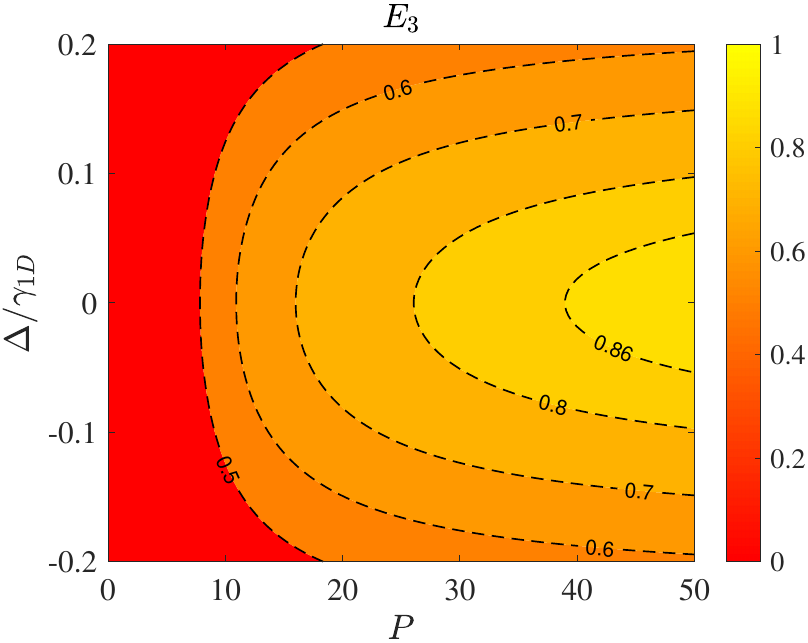}
			\label{fig11c}
	\end{minipage}}
	\subfigure[]
	{
		\begin{minipage}{0.75\linewidth}
			\centering
			\includegraphics[width=0.9\linewidth]{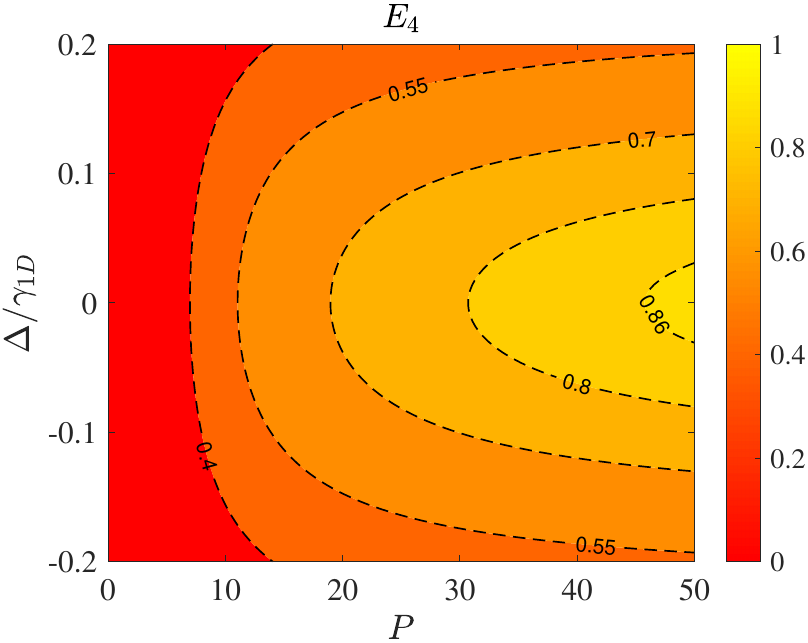}
			\label{fig11d}
	\end{minipage}}
	\caption{(a) The  efficiency $E_{1}$ of  the three-qudit entangled  $|\varphi_{000}\rangle$ (or $|\varphi_{011}\rangle$, $|\varphi_{022}\rangle$, $|\varphi_{033}\rangle$);  (b) the efficiency $E_{2}$ of the state  $|\varphi_{001}\rangle$ (or $|\varphi_{010}\rangle$, $|\varphi_{023}\rangle$,  $|\varphi_{032}\rangle$); (c) the efficiency $E_{3}$ of the state  $|\varphi_{002}\rangle$ (or $|\varphi_{013}\rangle$, $|\varphi_{020}\rangle$, $|\varphi_{031}\rangle$); (d) the efficiency $E_{4}$ of the state the state $|\varphi_{003}\rangle$ (or $|\varphi_{012}\rangle$, $|\varphi_{021}\rangle$, $|\varphi_{030}\rangle$) vs the Purcell factors $P$ and the frequency detuning $\Delta/{{\gamma }_{1D}}$.} \label{fig11}
\end{figure}

To sum up, our proposals  can also be extended to the $d$D $n$-qudit maximally entangled state, and that the number of emitters required is $\frac{1}{4}(n-1)*(4+d)$. The $d$D $n$-qudit entangle state
\begin{eqnarray}\label{eq37}
	\left|\phi_{00...0}\right\rangle=\frac{1}{\sqrt{d}} \sum_{l=0}^{d-1} |l\rangle_{12...n}^{\otimes n}
\end{eqnarray}
can be obtained. The fundamental principle of our protocol is rooted in the photon-emitter scattering dynamics within a 1D waveguide, so the HD $n$-qudits entanglement for hybrid system can be achieved by the error-detected mechanism
of the emitter-waveguide systems.

The reflection coefficient for photon-emitter scattering in a 1D waveguide system is given by $r=-1/(1+1/P-2i\Delta/\gamma_{1D})$, the imperfection in phase and amplitude of the output photon
reduces the performance of  qudit-based HD entanglement generation for hybrid
photon-emitter system, so it is necessary to consider
the feasibilities of  HD entangled states, which can be evaluated by the
fidelity defined as $F=|\langle\Phi_r|\Phi_i\rangle|^2$ \cite{hu2008giant,hu2008deterministic}, where $|\Phi_r\rangle$ and $|\Phi_i\rangle$ are the final states of the hybrid system after being processed in realistic and ideal situations, respectively. In the ideal case, the emitter-waveguide system
is ideal corresponding to the reflection coefficient $r=-1$.
In 2018,  Shailesh et al. \cite{kumar2018extremely} demonstrated that NV centers could achieve a Purcell factor of
$P=50$ by exciting plasmonic gap waveguide modes in silver nanowire-plate hybrid structures. More recently, Gritsch et al. \cite{gritsch2023purcell} reported a significant enhancement of
$P=78$ through the integration of erbium dopants into optimized nanophotonic silicon resonators under resonant excitation conditions. These advancements underscore the field's progression from fundamental mode engineering to sophisticated hybrid material architectures.
Taking the generation of 4D  two-qudit entangled states as an
example, where the ideal two-qudit entangled states can be described by in  Eq. (\ref{eq15}),
and the corresponding realistic output states shown in Eqs. (\ref{eq19})-(\ref{eq22}).
Four two-qudit entangled states, where  only the input ports are different and the interaction with two emitters is the same,
have the same fidelity,
therefore, we adopt the fidelity $F$ of the 4D two-qudit entangled state $|\Phi\rangle_{2}$
to display their features.
The fidelity $F$
varies with the frequency detuning $\Delta/\gamma_{1D}$ and the Purcell factor $P$ shown in Fig. \ref{fig8}(a), respectively, which enhancement scales significantly with increasing Purcell factor  $P$  for an established frequency detuning $\Delta/\gamma_{1D}$. Alternatively, for an established Purcell factor, the fidelity $F$ decreases with the increment of the frequency detuning $\Delta/\gamma_{1D}$. The fidelity is  $F$=99.97\% with $P=40$ and $\Delta /{{\gamma }_{1D}}=0$ \cite{PhysRevLett.101.113903,basso2019spectral,mittelstadt2021terahertz}, and $F$=98.23\% with $P=25$ and  $\Delta /{{\gamma }_{1D}}=0.05$.

Efficiency is the other powerful and practical indicator for evaluating the performance of the HD entanglement generation.
The efficiency E for the heralding emitter-waveguide coupled system is quantified by the squared overlap integral between the input photon's spatial wavefunction $|\phi\rangle$ and the reflected photon's wavefunction $|\phi_r\rangle$, expressed as $E=\left|\langle\phi|\phi_r\rangle\right|^2$, consistent with the above theoretical framework. Fig. \ref{fig8}(b) shows the efficiency $E$ of the 4D two-qudit entangled state  $|\varphi_{00}\rangle$ (or $|\varphi_{01}\rangle$, $|\varphi_{02}\rangle$, and $|\varphi_{03}\rangle$) varies as a function of the frequency detuning $\Delta/\gamma_{1D}$ and the Purcell factor $P$, respectively. Evidently, for a stationary Purcell factor, the efficiency decreases with the frequency detuning $\Delta/\gamma_{1D}$, and the efficiency increase with the  Purcell factor $P$ for a stationary frequency detuning $\Delta/\gamma_{1D}$.
The efficiency  of the 4D two-qudit entangled state is $E$=95.21\% in case of $P=40$ and $\Delta /{{\gamma }_{1D}}=0$, and $E$=91.70\% in other case $P=25$ and  $\Delta /{{\gamma }_{1D}}=0.05$.
The rest two-qudit entangled states in  Eq. (\ref{eq16}) can be implemented through corresponding Pauli $Z^{1}$ gate, $Z^2$ gate, and  $Z^3$ gate, respectively.
As three Pauli $Z$ gates can be realized with linear optical element, i.e., three wave plates combination, in principle,
their fidelities and efficiencies are near to 100\%.
Therefore, we believe that  the fidelities and efficiencies of	the rest two-qudit entangled states in  Eq. (\ref{eq16}) are the same as
that of the two-qudit entangled state in  Eq. (\ref{eq15}) shown in Fig. \ref{fig8}.

Moreover, the fidelities and efficiency of 4D three-qudit X gates are shown in Figs. \ref{fig9}(a)-(c), respectively. The fidelity $F_{X}$($F_{X^{\dagger}}$) of the single-qudit $X$ ($X^{\dagger}$)  gate is shown in Fig. \ref{fig9}(a). In the condition $P=40$ and $\Delta /{{\gamma }_{1D}}=0$, the fidelity  is $F_{X}(F_{X^{\dagger}})=99.98\%$. Additional, the fidelity $F_{X^{2}}$ of $X^{2}$ gate approaches unity under ideal conditions.
The efficiency  $E_{X}$($E_{X^{\dagger}}$) of the single-qudit $X$ ($X^{\dagger}$) gate is shown in Fig. \ref{fig9}(b), the efficiency $E_{X^{2}}$  of the single-qudit $X^{2}$ gate is shown in Fig. \ref{fig9}(c). In the condition $P=40$ and $\Delta /{{\gamma }_{1D}}=0$, the efficiencies are  $E_{X}$($E_{X^{\dagger}})=92.85\%$, $E_{X^{2}}=95.27\%$.

Furthermore, the fidelities of  4D three-qudit entangled states are shown in Figs. \ref{fig10}(a)-(c). The fidelity $F_{1}$ of the three-qudit entangled state  $|\varphi_{000}\rangle$ (or $|\varphi_{011}\rangle$, $|\varphi_{022}\rangle$, $|\varphi_{033}\rangle$, $|\varphi_{002}\rangle$, $|\varphi_{013}\rangle$, $|\varphi_{020}\rangle$,  $|\varphi_{031}\rangle$) is shown in Fig. \ref{fig10}(a), the fidelity $F_{2}$ of the state  $|\varphi_{001}\rangle$ (or $|\varphi_{010}\rangle$, $|\varphi_{023}\rangle$,  $|\varphi_{032}\rangle$) is shown in Fig. \ref{fig10}(b), and the fidelity $F_{3}$ of the state $|\varphi_{003}\rangle$ (or $|\varphi_{012}\rangle$, $|\varphi_{021}\rangle$, $|\varphi_{030}\rangle$) is shown in Fig. \ref{fig10}(c).
In the condition $P=40$ and $\Delta /{{\gamma }_{1D}}=0$, the fidelities are $F_{1}=99.87\%$, $F_{2}=99.79\%$, and $F_{3}=99.92\%$, respectively.
The efficiencies of  4D three-qudit entangled states are shown in Figs. \ref{fig11}(a)-(d).
The efficiency $E_{1}$ of the state  $|\varphi_{000}\rangle$ (or $|\varphi_{011}\rangle$, $|\varphi_{022}\rangle$, $|\varphi_{033}\rangle$) is shown in Fig. \ref{fig11}(a),
the efficiency $E_{2}$ of the state  $|\varphi_{001}\rangle$ (or $|\varphi_{010}\rangle$, $|\varphi_{023}\rangle$,  $|\varphi_{032}\rangle$) is shown in Fig. \ref{fig11}(b),
the efficiency $E_{3}$ of the state  $|\varphi_{002}\rangle$ (or $|\varphi_{013}\rangle$, $|\varphi_{020}\rangle$, $|\varphi_{031}\rangle$) is shown in Fig. \ref{fig11}(c),
and	the efficiency $E_{4}$ of the state the state $|\varphi_{003}\rangle$ (or $|\varphi_{012}\rangle$, $|\varphi_{021}\rangle$, $|\varphi_{030}\rangle$) is shown in Fig. \ref{fig11}(d).
In the condition $P=40$ and $\Delta /{{\gamma }_{1D}}=0$, the efficiencies are $E_{1}=90.71\%$, $E_{2}=84.29\%$,
$E_{3}=86.33\%$
and $E_{4}=84.29\%$, respectively.
Obviously, the fidelities  and efficiencies of these three-qudit entangled states increase
by reducing  the frequency detuning $\Delta/\gamma_{1D}$ and scaling up the Purcell factor $P$.

\section{Summary}\label{sec7}

In summary, we have investigated the flexible generation of random 4D  two-qudit maximally entangled states for hybrid photon-emitter system according to different input ports, resorting to the
error-detected mechanism of the emitter-waveguide systems, and then it can be extended the generation of random 4D $n$-qudit ($n\geq3$) maximally entangled states with the $2n-2$ photon-emitters systems by virtue of the 4D single-qudit $Z^{m} (m=1,2,3)$ gate for the first qudit and $X^{m}$ gate for the other qudits (except the second qudit). Moreover, the proposed protocol can be spread to generate $d$D
$n$-qudit ($d\geq 2^{p+1}$, $n, p=2,3,\dots$) entangled states, takeing generation of random 8D  two-qudit maximally entangled states as an example.
 The proposed protocols of HD entanglement not only save quantum resources without any auxiliary qudits but also enhance resistance to loss. Remarkably, the frequency detuning between the photon and emitter can be precisely controlled. This technological capability suggests that our protocols may soon be implemented with robust fidelities and exceptional efficiencies as photonic technologies continue to advance. As the emitter-waveguide system advances and error-detected mechanism is introduced, our schemes for generation of HD entangled states will be feasible experimentally.

\begin{acknowledgments}
	This work was supported in part by the Natural Science Foundation of China under Contract 61901420;	
	in part by Fundamental Research Program of Shanxi Province under Contract 20230302121116.
	
	The authors declare no conflicts of interest.
\end{acknowledgments}

\section*{Data Availability}
The data supporting this study findings are available
within the article.


\bibliography{PRAreferences}

\end{document}